%% file: fomalhaut.tex
\documentclass[longauth]{aa} 
\usepackage{graphicx}
\usepackage{color}
\usepackage{epsfig}

\usepackage{amssymb, amsmath}
\usepackage{graphicx,natbib}
\usepackage{pdflscape}

\usepackage{longtable}
\usepackage{lscape}
\usepackage[figuresright]{rotating} 

\usepackage{url}
\usepackage{textcomp}

\usepackage{multirow}
\usepackage{upgreek}

\input{commands}
\begin{document}
\authorrunning{J. Lebreton et al.}
\title{An interferometric study of the Fomalhaut inner debris disk}
\subtitle{III. Detailed models of the exozodiacal disk and its origin}
\titlerunning{The Fomalhaut inner debris disk}
\author{
     J. Lebreton\inst{1}, R. van Lieshout\inst{2}, J.-C. Augereau\inst{1}, O. Absil\inst{3}, B. Mennesson\inst{4}, M. Kama\inst{2}, C. Dominik\inst{2,5}, A.~Bonsor\inst{1}, J. Vandeportal\inst{1,6}, H. Beust\inst{1}, D. Defr\`ere\inst{7}, S. Ertel\inst{1}, V. Faramaz\inst{1}, P. Hinz\inst{7}, Q. Kral\inst{8}, \\
     A.-M. Lagrange\inst{1}, W.~Liu\inst{9}, and P. Th\'ebault\inst{8}
}
\institute{
UJF-Grenoble 1 / CNRS-INSU, Institut de Plan\'etologie et d'Astrophysique de Grenoble (IPAG) UMR 5274, Grenoble, F-38041, France
\and Astronomical Institute ``Anton Pannekoek'', University of Amsterdam, P.O. Box 94249, 1090 GE Amsterdam, The Netherlands
\and Institut d'Astrophysique et de G\'eophysique, Universit\'e de Li\`ege, 17, All\'ee du 6 Ao\^ut, B5c B-4000 Sart Tilman, Belgium
\and Jet Propulsion Laboratory, California Institute of Technology, Pasadena, CA 91109, USA
\and Department of Astrophysics/IMAPP, Radboud University Nijmegen, P.O. Box 9010, 6500 GL Nijmegen, The Netherlands
\and D\'epartement de Physique and Observatoire du Mont-M\'egantic, Universit\'e de Montr\'eal, C.P. 6128, Succ. Centre-Ville, Montr\'eal, Qu\'ebec H3C 3J7, Canada
\and Steward Observatory, University of Arizona, 933 N Cherry Avenue, Tucson, AZ 85721, USA 
\and Observatoire de Paris, Section de Meudon, F-92195 Meudon Principal Cedex, France
\and Infrared Processing and Analysis Center, California Institute of Technology, Mail Code 100-22, Pasadena, CA 91125, USA
}
\offprints{jeremy.lebreton@obs.ujf-grenoble.fr}
\date{Received March 5, 2013; accepted May 31, 2013}
\abstract{Debris disks are thought to be extrasolar analogs to the solar system planetesimal belts. The star Fomalhaut harbors a cold debris belt at 140\,AU comparable to the Edgeworth-Kuiper belt, as well as evidence of a warm dust component, unresolved by single-dish telescopes, which is suspected of being a bright analog to the solar system's zodiacal dust.}
{Interferometric observations obtained with the VLTI/VINCI instrument and the Keck Interferometer Nuller have identified near- and mid-infrared excesses attributed respectively to hot and warm exozodiacal dust residing in the inner few AU of the Fomalhaut environment. We aim to characterize the properties of this double inner dust belt and to unveil its origin.}
{We performed parametric modeling of the exozodiacal disk (``exozodi") using the GRaTeR radiative transfer code to reproduce the interferometric data, complemented by mid- to far-infrared photometric measurements from \textit{Spitzer} and \textit{Herschel}\thanks{\textit{Herschel} Space Observatory is an ESA space observatory with science instruments provided by European-led Principal Investigator consortia and with important participation from NASA.}. A detailed treatment of sublimation temperatures was introduced to explore the hot population at the size-dependent sublimation rim. We then used an analytical approach to successively testing several source mechanisms for the dust and suspected parent bodies.}
{A good fit to the multiwavelength data is found by two distinct dust populations: (1) a population of very small (0.01 to 0.5 $\mu$m), hence unbound, hot dust grains confined in a narrow region ($\sim$0.1 -- 0.3 AU) at the sublimation rim of carbonaceous material; (2) a population of bound grains at $\sim2$AU that is protected from sublimation and has a higher mass despite its fainter flux level. We propose that the hot dust is produced by the release of small carbon grains following the disruption of dust aggregates that originate from the warm component. A mechanism, such as gas braking, is required to further confine the small grains for a long enough time. \textit{In situ} dust production could hardly be ensured for the age of the star, so we conclude that the observed amount of dust is triggered by intense dynamical activity.}
{Fomalhaut may be representative of exozodis that are currently being surveyed at near and mid-infrared wavelengths worldwide. 
We propose a framework for reconciling the ``hot exozodi phenomenon" with theoretical constraints: the hot component of Fomalhaut is likely the ``tip of the iceberg" since it could originate from the more massive, but fainter, warm dust component residing near the ice line. 
This inner disk exhibits interesting morphology and can be considered a prime target for future exoplanet research.}
\maketitle
%
\input{fom_text}
\input{rik5_sec5}

\input{conclusion}

\begin{acknowledgements}
We would like to acknowledge Paul Kalas, James Graham, Kate Su and Alexis Brandeker, for contributing to interesting discussions on various aspects of this study, as well as the anonymous referee for the valuable advises he provided.
The research leading to these results has received funding from the European Community's Seventh Framework Programme under Grant Agreement 226604.
We also thank the French National Research Agency
(ANR) for financial support through contract ANR-2010 BLAN-0505-01
(EXOZODI), the Programme
National de Plan\'etologie (PNP) and the CNES for supporting part of
this research. 
\end{acknowledgements}
\bibliography{biblio,bibbli_rik}
\appendix
\input{rik4_app_black}

\section{Bayesian probability curves}
\input{bayes}

\end{document}

%% file: commands.tex
\newcommand{\uma}[1]{^{\mathrm{#1}}}
\newcommand{\dma}[1]{_{\mathrm{#1}}}

\newcommand{\dd}{\textrm{d}}
\def\@object[#1]#2{#2}%
\def\q1{\object{q1~Eri}}
\def\bp{\object{$\beta$~Pictoris}}
\def\fom{\object{Fomalhaut}}

\def\d{\mathrm d}

\DeclareMathAlphabet{\mathpzc}{OT1}{pzc}{m}{it}

\DeclareTextSymbol{\degre}{OT1}{23}
\newcommand{\earth}{\mathrm{\oplus}}


%% file: fom_text.tex
\section{Introduction}
During the past few years, the increasing number of smaller exoplanets and fainter debris disks have revealed that extrasolar analogs to our solar system may be common, and yet, little is known about the architecture of the very inner parts of planetary systems. A distinguishable feature of the inner solar system is the existence of the zodiacal cloud, composed of small (1 to 100 $\mu$m) dust grains \citep{Grun2007621, 2013MNRAS.tmp..533R}, which are thought to come from the disruption and erosion of comets, asteroids, and Kuiper belt objects \citep[\textit{e.g.}][]{2010ApJ...713..816N}. 
Dusty debris disks orbiting other stars than the Sun were first detected by their excess emission in the mid- or far-infrared (IR), and could then be imaged at visible to submillimeter wavelengths. A few warm disks comparable to the zodiacal cloud have been found by space observatories around mature stars via their photometric excess emission at mid-IR wavelengths \citep{2005ApJ...626.1061B,2009ApJ...705...89L,2012ApJ...747...93L}; but their characterization suffers from insufficient spatial resolution and large photometric uncertainties. 

Recent developments in high-angular resolution interferometry have offered powerful tools to characterizing exozodiacal disks (exozodis), which reside in the close environment (typically less than 3 AU) of a large fraction of nearby stars. Large efforts have indeed been made to detect exozodis with near- and mid-IR interferometers worldwide, notably at the VLTI (Absil et al. 2009), the Keck Interferometer \citep{2011ApJ...734...67M} or the CHARA array at Mount-Wilson \citep{2006A&A...452..237A,2011epsc.conf.1084D}. Ongoing surveys in the near-infrared (K $<$ 5) with the CHARA/FLUOR \citep{2003SPIE.4838..280C} and VLTI/PIONIER \citep{2011A&A...535A..67L} interferometers indicate that as much as $\sim\,30\%$ of nearby AFGK mainsequence stars may harbor hot (typically 1000-2000K) circumstellar dust within a few AU at the 1\% level with respect to photospheric emission (Absil et al., in prep., Ertel et al., in prep.). Conversely, only $\sim12\%$ of the surveyed main sequence stars have been found to harbor mid-infrared excesses (i.e., warm dust) with nulling interferometry \citep{2011ApJ...734...67M}.

From a theoretical point of view, the prevalence of these hot excesses around main sequence stars is not understood. Radiative transfer analysis identifies very hot and small refractory grains close to the sublimation limit that should be radiatively blown out over timescales of weeks. Yet they represent typical masses of $10^{-8}$ to $10^{-10}$ $M\dma{\oplus}$ that need to be delivered by equivalent masses of dust-producing planetesimals \citep[\textit{e.g.}][]{2011A&A...534A...5D}. 
Being much more massive than the zodiacal cloud, these hot exozodis are difficult to reconcile with the steady-state collisional evolution of parent body belts (Wyatt 2007). 

One famous example of a dusty planetary system is the one surrounding the nearby (7.7 pc) A3V star Fomalhaut ($\alpha$ PsA, {HD\,216956}). This young main sequence star \citep[$440\pm40$ Myr,][]{2012ApJ...754L..20M} is well known for its prominent, $\sim140$\,AU-wide, debris belt that was first resolved in scattered light with HST/ACS revealing a sharp inner edge and side-to-side brightness asymmetry suggestive of gravitational sculpting by a massive planet \citep{2005Natur.435.1067K,2006MNRAS.372L..14Q}. 
A point source attributed to the suspected planet was later detected in the optical at the predicted location, moving along its orbit over multiple epochs \citep{2008Sci...322.1345K}. The absence of detection in the near- and mid- (thermal) infrared range \citep{2012ApJ...747..116J} resulted in a controversial status for Fomalhaut b. 
More recent studies confirm the detections of a companion at 118\,AU and interpret the various constraints as a large circumplanetary dust disk orbiting a hidden subjovian planet, but do not exclude an isolated dust cloud originating from a recent collision between planetesimals \citep{2012arXiv1210.6745G}\footnote{\citet{2013arXiv1305.2222K} recently announced a fourth epoch detection of Fomalhaut b. They argue in favor of an object larger than a dwarf planet evolving on a very eccentric orbit (see Section 6).}.
The cold planetesimal belt is indeed collisionally very active as shown by the recent analysis of resolved images and photometry at far-infrared to submillimeter wavelengths from ALMA \citep{2012ApJ...750L..21B} and \textit{Herschel}/PACS and SPIRE \citep{2012A&A...540A.125A}. 
Using radiative transfer modeling, \citet{2012A&A...540A.125A} argue that a dust production rate of $\sim\,3\times 10^{-5}\,M\dma{\oplus}$/year is needed to justify the measured amount of blow-out grains, and estimate that the region interior to the cold belt at 140\,AU are not devoid of material.

The \textit{Herschel}/PACS 70\,$\mu$m, as well as the ALMA image, identify an unresolved excess in the vicinity of the star that was previously reported in the mid-infrared by \citet{2004ApJS..154..458S} based on \textit{Spitzer}/MIPS imaging and IRS spectroscopy. \citet{2013ApJ...763..118S} recently analyzed the IRS and PACS data and concluded on the presence of a warm debris belt with a blackbody temperature of $\sim170$\,K.
However, these facilities lack the spatial resolution and accuracy needed to characterize this warm component.

Near- and mid-infrared long baseline interferometers offer the appropriate tools to study the Fomalhaut exozodi with enough contrast and resolution, free of any modeling assumptions on the stellar spectrum.

The present paper carries out a thorough analysis of the Fomalhaut inner debris disk. It is the last one of a series initiated by \citet{2009ApJ...704..150A} (henceforth Paper I) - who presented the VLTI/VINCI detection of hot excess in the close environment of the star - followed by a mid-infrared characterization using the Keck Interferometer Nuller (KIN) by \citet{2013ApJ...763..119M} (henceforth Paper II).
In Paper I, we presented the clear K-band (2.18$\mu$m) detection of a short-baseline visibility deficit. 
It is best explained by circumstellar emission emanating from within 6 AU of the star with a relative flux level of $0.88\pm 0.12\%$.
In Paper II, we presented multiwavelength measurements performed across the N-band (8 to 13 $\mu$m) using the technique of nulling interferometry. 
A small excess is resolved within $\sim\,2$AU from the star, with a mean null depth value of $0.35\% \pm 0.10\%$ \footnote{We note that the true astrophysical excess could be larger because some dust emission may be removed by the destructive fringes of the interferometer. The nuller transmission pattern is accounted for in the models discussed afterwards.}. Preliminary modeling shows that the near- to mid-infrared excesses can only be explained by two distinct populations of dust emitting thermally; small ($\sim$20 nm) refractory grains residing at the sublimation distance of carbon are responsible for the near-infrared emission, while $> 1\,\mu$m grains located further than the silicate sublimation limit produce most of the mid-infrared excesses.

The paper is organized as follows. In Sec.\,\ref{sec:grater} and \ref{sec:sublimation}, we present our radiative transfer model of optically thin disks, and introduce a new prescription for calculating the sublimation distance of dust grains in an exozodi. Results of the modeling of the inner Fomalhaut debris disk, based on multiwavelengths observations, are detailed in Sec.\ref{sec:results}. The mechanisms that produce and preserve the hot grains, as well as the connection with the warm and the cold belt are discussed in Sec.\,\ref{sec:origin}. 
We discuss further our results and attempt to place the Fomalhaut exozodi in the context of its planetary system in Sec.\,\ref{sec:discu}. 
We finally summarize our main findings in Sec.\ref{sec:conclu}.

\section{A schematic exozodiacal disk model}\label{sec:grater}
In this section, we elaborate a schematic model of an optically thin
exozodiacal dust disk. The model is implemented in the GRaTeR code originally developped by \citet{Augereau1999a}.
It makes hardly any \textit{a priori} assumptions regarding the
nature of the grains and their production. 
Because of the limited constraints on detected exozodiacal disks, we restrain
the model to a 2D geometry. The disk surface density and grain size
distribution are parametrized with simple laws to limit the number of
free parameters. This allows us to explore a broad range of disk
properties using a Bayesian inference method.

\subsection{Scattered light and thermal emission}
We consider a population of dust grains at a distance $r$ from the
star and with a differential size distribution $\d n(r,a)$, where $a$
is the grain radius. In a self-consistent description of a collisional
evolution of debris disks, the spatial and size distribution cannot be
formally separated \citep{2001A&A...370..447A, 2006A&A...455..509K,2007A&A...472..169T}. Since we know little
about the properties and origin of observed exozodiacal disks, we
assume here for simplicity that the dependence of the size
distribution on the distance $r$ essentially reflects the
size-dependent sublimation distance of the grains. Since exozodiacal
grains may reach very high temperatures, sublimation can prevent the
smallest grains from surviving in regions where larger ones can remain,
thereby truncating the size distribution at its lower end. We
therefore write the differential size distribution $\d n(r,a)$ as
follows:
\begin{equation}
\begin{array}{ll}
  \d n(r,a) = H(a-a\dma{sub}(r))\,\, \d n(a) \\
  
  \mathrm{\,\,\, with \,}
  \int_{a\dma{min}}^{a\dma{max}}\d n(a) = 1
\end{array}
\end{equation}
where $a\dma{min}$ and $a\dma{max}$ are the minimum and maximum grain
sizes respectively, $H(a-a\dma{sub}(r))$ is the Heaviside function
(assuming $H(0)=1$) and $a\dma{sub}(r)$ is the sublimation size limit
at distance $r$ from the star. Details of the calculation for sublimation will be given in Sec.\,\ref{sec:sublimation}. 

At wavelength $\lambda$, the dust population thermally emits a flux
\begin{eqnarray}
  \Phi\dma{th}(\lambda,r) & = & \int_{a\dma{min}}^{a\dma{max}}
  B_{\lambda}\left(T\dma{d}(a,r)\right)
  \frac{\sigma\dma{abs}(\lambda,r,a)}{4 d_\star^2} \d n(r,a)
\end{eqnarray}
where $d_\star$ is the distance of the observer to the star,
$T\dma{d}(a,r)$ is the grain temperature and $B_{\lambda}$ is the Planck function. 
In the above equation, we implicitly assumed that grains thermally emit isotropically. The
absorption cross-section $\sigma\dma{abs}(\lambda,r,a)$ reads
\begin{eqnarray}
  \sigma\dma{abs}(\lambda,r,a) & =  & 4\pi a^2  
  Q\dma{abs}\left(\frac{2\pi a}{\lambda},\lambda,r\right)
\label{th}
\end{eqnarray}
where $Q\dma{abs}$ is dimensionless absorption/emission coefficient
that depends on the size parameter $2\pi a/\lambda$, on $\lambda$
through the wavelength-dependent optical constants, and on the
distance to the star as the grain composition may depend on $r$. The
grain temperature $T\dma{d}(a,r)$ is obtained by solving in two steps
the thermal equilibrium equation of a grain with the star. For any
grain size $a$, we first calculate the equilibrium distance $r$ for a broad range
of grain temperatures $T\dma{d}$ knowing the $Q\dma{abs}$ value
\begin{eqnarray}
  r(a,T\dma{d}) = \frac{d_\star}{2}\sqrt{\frac{
      \int_{\lambda} Q\dma{abs}F_{\star}(\lambda)\d \lambda}
    {\int_{\lambda} Q\dma{abs} \pi B_{\lambda}\left(T\dma{d}\right)\d
      \lambda}}
  \label{eq:Tdust}
\end{eqnarray}
where $F_{\star}(\lambda)$ is the stellar flux at Earth. The
$r(a,T\dma{d})$ function is then numerically reversed to get
$T\dma{d}(a,r)$. 

Assuming isotropic scattering for simplicity, a dust population at
distance $r$ from the star scatters a fraction of the stellar flux at
wavelength $\lambda$ is given by
\begin{eqnarray}
  \Phi\dma{sc}(\lambda,r) & = & 
  F_\star(\lambda) \frac{\sigma\dma{sca}(\lambda,r)}{4\pi r^2} 
\end{eqnarray}
with $\sigma\dma{sca}$ the mean scattering cross section
\begin{eqnarray}
  \sigma\dma{sca}(\lambda,r) & = & 
  \int_{a\dma{min}}^{a\dma{max}}\pi a^2  
  Q\dma{sca}\left(\frac{2\pi a}{\lambda},\lambda,r\right)\d n(r,a)
\end{eqnarray}
and $Q\dma{sca}$ the dimensionless scattering coefficient.

The total flux emitted at wavelength $\lambda$ by the dust population
finally reads
\begin{eqnarray}
\Phi(\lambda,r) & = & \Phi\dma{sc}(\lambda,r) + \Phi\dma{th}(\lambda,r).
\end{eqnarray}

\subsection{Synthetic observations}
We synthesize single aperture photometric observations of non edge-on
exozodiacal dust disks at wavelength $\lambda$ by calculating the
integral
\begin{eqnarray}
  \Phi(\lambda) & = & \int_{r=0}^{\infty} \Phi(\lambda,r)
  \Sigma(r) \times 2\pi\langle FOV\rangle_{\theta}(r)  r \d r \label{eq:bbflux}\\
  \mathrm{with} & & \langle FOV\rangle_{\theta}(r) = 
  \int_{\theta=0}^{2\pi} FOV(\rho(r,\theta)) \frac{\d \theta}{2\pi} \\
  \mathrm{and} & &  \rho(r,\theta) = r \sqrt{1-\cos^2 {\theta}\sin^2{i}}
\end{eqnarray}
where $r$ and $\theta$ are cylindrical coordinates in the disk plane
and $\rho(r,\theta)$ the projected distance to the star in the sky
plane, $\Sigma(r)$ is the dust surface number density of the
exozodiacal disk, assumed to be axisymmetrical and $i$ is the disk
inclination with respect to the sky plane ($i=0$ for pole-on
geometry). In the above equations, we implicitly assumed that the
instrument beam profile $FOV$ only depends upon $\rho$, the projected
distance to the star in the sky plane. The $\langle
FOV\rangle_{\theta}(r)$ function gives the azimuthally averaged
telescope transmission along circles of radius $r$ in the exozodiacal
disk frame.

Most single aperture telescopes such as \textit{Spitzer} or \textit{Herschel},
have sufficiently large beams (or slits in case of spectroscopy), which
intercept the entire exozodiacal dust emission. In such a case, the
exozodiacal flux emission can be obtained by taking $\langle
FOV\rangle_{\theta}(r) =1$ in Eq.\,\ref{eq:bbflux}. On the other hand,
coherent near- and mid-IR interferometric observations, such as those
obtained with the VLTI/VINCI and CHARA/FLUOR instruments, have much
smaller fields of view (FOV) and the actual transmission profiles of the
interferometric instruments on the sky plane is important\footnote{See Fig.\,1 of Paper II for the KIN transmission map}.

\subsection{Grain properties}
In the solar system, the zodiacal dust particles are thought to
originate from tails and disruption of comets, or to be produced when asteroids
collide. Both interplanetary and cometary dust particles are composed of silicates and carbonaceous material,
and zodiacal cloud dust particles are expected to be made of similar material. In this model,
we consider mixtures of silicates and carbonaceous material, and
calculate their optical properties with the Mie theory valid for hard
spheres. The optical index of the mixture is calculated using the
Bruggeman mixing rule, given a relative volume fraction $v\dma{C}/v\dma{Si}$ of carbonaceous grains. In the disk regions where the grain temperature is
between the sublimation temperature of silicates and carbon,
the volume of silicates is replaced by vacuum, mimicking porous carbonaceous grains.
The optical constants and grain bulk densities used in this study are summarized in Tab.\,\ref{tab:parameters}.

We adopt a power-law differential size distribution
\begin{eqnarray}
  \d n(a) & = & 
  \left(\frac{1-\kappa}{a\dma{max}^{1-\kappa} - a\dma{min}^{1-\kappa}}\right)
  a^{-\kappa}\d a
\end{eqnarray}
for grain radii $a$ between $a\dma{min}$ and $a\dma{max}$. The maximum
grain size has no impact at the wavelengths considered and 
cannot be constrained with the adopted modeling approach. It is thus fixed to $a\dma{max}=1$\,mm in the rest of this study.

\subsection{Fitting strategy}

The adopted fitting stategy is based on a Bayesian inference
method described in \citet{Lebreton2012a}. For that purpose, a grid of models is created. For each set of
parameters, the goodness of the fit is evalutated with a reduced
$\chi\dma{r}^2$ that is transformed into probabilities assuming a
Gaussian likelihood function 
($\propto e^{-\chi^2\dma{r} /2}$)
for
Bayesian analysis. Marginalized probability distributions for each free
parameter are then obtained by projection of these probabilities onto
each dimension of the parameter space.

In order to limit the number of free parameters, we adopt a two
power-law radial profile for the surface density
\begin{eqnarray}
  \Sigma(r) = \Sigma_0 \sqrt{2}\left[\left(\frac{r}{r_0}\right)^{-2\alpha\dma{in}} + 
      \left(\frac{r}{r_0}\right)^{-2\alpha\dma{out}} \right]^{-1/2}.
\end{eqnarray}
This corresponds to a smooth profile, with an inner slope $r^{\alpha\dma{in}}$ ($\alpha\dma{in} >
0$) that peaks at about $r_0$, and that falls off as $r^{\alpha\dma{out}}$ ($\alpha\dma{out} <
0$) further to $r_0$.

The parameter space explored here is listed in
Tab.\,4 of Paper II and recalled in Sec.\,\ref{sec:model_st}. Each set of parameters defines an emission spectrum, 2D
emission, and scattered light maps that are flux-scaled by searching
the optimal $\Sigma_0$ value (surface density at $r=r_0$) that gives
the best fit to the panchromatic observations. The range of
$a\dma{min}$ values includes sizes that are far below the blow-out
size limit for grains about \fom.
\section{Dust sublimation model}\label{sec:sublimation}
\subsection{Sublimation temperatures}
We implement a new method of calculating the sublimation temperature $T\dma{sub}$ of a dust grain as a function of its size and not only of its composition, based on the method introduced by \citet{2009A&A...506.1199K} to study the inner rim of protoplanetary disks.
We first model a grain as an homogeneous sphere of radius $a$ composed of a single material of density $\rho\dma{{d}}$. 
To ensure the stability of a grain (\textit{i.e.} no net change of its size), the flux of particles (carbon / silicates monomeres) escaping from its surface must equal the flux of particles coming in from the ambient medium. 
From the kinetic theory of gases, the number of particles accreted onto and evaporated from the grain surface per unit time and unit surface reads respectively \citep{1974A&A....35..197L}

\begin{eqnarray}
	F\dma{accre}&=&\dfrac{P\dma{gas}}{\sqrt{2\pi\mu\,m\dma{u}\,k\dma{B}\,T\dma{\textrm{gas}}}}\\
	F\dma{evap}&=&-\dfrac{P\dma{eq}}{\sqrt{2\pi\mu\,m\dma{u}\,k\dma{B}\,T\dma{\textrm{gas}}}}
\end{eqnarray}
where $P\dma{eq}$ is the gas saturation partial pressure, $P\dma{gas}$ the partial pressure of the ambient gas medium, $T\dma{\textrm{gas}}$ the gas temperature assumed equal to the dust temperature $T\dma{\textrm{d}}$ here, $\mu m\dma{u}$ the mean molecular weight and $k\dma{B}$ the Boltzmann constant. Introducing an efficiency factor $\alpha$ (constrained by laboratory experiments), the mass of a grains of radius $a$ then evolves as

\begin{equation}
\dfrac{\dd m}{\dd t} = \alpha\,( F\dma{accre} - F\dma{evap} ) \times \mu\,m\dma{u}\, 4\,\pi a^2
\end{equation}

Injecting the ideal gas law $P\,=\,\rho k\dma{B} T/ \mu m\dma{u}$ yields:
\begin{eqnarray}
\dfrac{\dd m}{\dd t}&=&\alpha\,a^2\,\sqrt{\frac{8\pi\,k\dma{B}\,T\dma{{d}}}{\mu\,m\dma{u}}}\, (\rho\dma{gas}-\rho\dma{eq})\\
\dfrac{\dd a}{\dd t}&=&\frac{\alpha}{\rho\dma{d}} \sqrt{\frac{k\dma{B}\,T\dma{d}}{2\pi\,\mu\,m\dma{u}}}\, (\rho\dma{gas}-\rho\dma{eq}).\label{eq:asub}
\end{eqnarray}

For the purpose of this study, the grains are assumed to lie in empty space ($\rho\dma{gas}=0$) and their temperature equals by definition the sublimation temperature $T\dma{sub}$. 
The gas density at saturation pressure $\rho\dma{eq}$ is given by the Clausius-Clapeyron equation
\begin{equation}
\log_{10}{\rho\dma{eq}} = B - \dfrac{A}{T\dma{sub}} - \log_{10}{T\dma{sub}\uma{-C}}
\end{equation}
where the thermodynamical quantities $A$ and $B$ are determined from laboratory measurements and $C = -1$, as discussed by \citet{2009A&A...506.1199K}.
Integrating equation \ref{eq:asub} from the initial grain size $a$ to 0, under the assumption that the grain temperature $T\dma{d} = T\dma{sub}$ does not vary significantly during the sublimation process, yields

\begin{equation}\label{eq:Tsub}
\dfrac{a}{t\dma{sub}} = {\frac{\alpha}{\rho\dma{d}}} \sqrt{\dfrac{k\dma{B}\,T\dma{sub}}{2\pi\,\mu m\dma{u}}} 10\uma{B-\frac{A}{T\dma{sub}}-log_{10}{T\dma{sub}}}
\end{equation}
where we define $t\dma{sub}$ as the time needed to sublimate an entire grain.

This equation relates the sublimation temperature of a grain to its size and to a sublimation timescale.
Although determining this timescale without a time-dependent approach is tricky, we will see in Sec.\,\ref{sec:timescales} that our parametric approach allows to tackle the issue based on simple assumptions. Beforehand we need to extend the sublimation model to the case of multi-material grains. \\
%
We recall that our objective is to interpret real observations of individual debris disks. 
Fitting the spectral and spatial observables of these disks requires to solve accurately a radiative transfer problem and to model the optical properties of the materials at stake. 
To achieve this, the GRaTeR code handles grains made of multiple (in particular carbonaceous and silicate) materials by considering homogeneous spheres with optical constants obtained by means of an effective medium theory. Porous grains are represented by a compact sphere in which one of the ``materials" is vacuum. When one of the material reaches its sublimation temperature, it is automatically replaced by vacuum. This is in particular the case for the grains that lie very close to the star: then a silicate-carbon mixture becomes a porous carbon grain. 
This approach proved efficient at reproducing the optical behavior of astronomical dust grains in various situations.

We consider homogeneously distributed mixtures of silicate and carbon characterized by a volume fraction $v\dma{C}/v\dma{Si}$. 
We anticipate that silicate will sublimate at lower temperatures than carbon whatever the grain size: silicates will have vanished entirely before the carbon starts its sublimation. 
As a first step the sublimation occurs as if the grains were entirely made of silicates.
In a second step, the grains take the form of a carbon matrix filled with cavities. 
For spheroids, the porosity can be defined as the filling factor of vacuum $\mathcal{P} = \frac{V\dma{vacuum}}{V\dma{grain}}$. We introduce the porosity by correcting the grain density:
$\rho\dma{d}(\mathcal{P}) \rightarrow \left({1-\mathcal{P}}\right){\rho\dma{d}}$ in 
Eq.\,\ref{eq:Tsub} which is then solved using the density of the porous carbonaceous leftover.

\begin{figure}[h!]\begin{center}   
  \includegraphics[angle=0,width=1.\columnwidth,origin=bl]{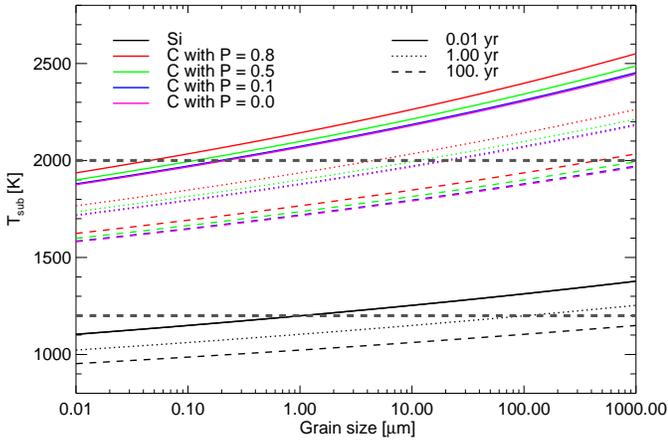}
     \caption{Sublimation temperatures of carbon and silicate as a function of grain size with several possible survival timescales, and assuming different porosities for carbonaceous grains or equivalently different volume fractions of silicate. 
     The horizontal dashed lines show the constant $T\dma{sub}$ values often used in past studies.}\label{fig:Tsub} 
 \end{center}\end{figure}

Eq.\,\ref{eq:Tsub} is solved using the material properties summarized in Tab.\,\ref{tab:parameters}, and the solution is inversed numerically to determine $T\dma{sub}$ as a function of grain size for different timescales. The sublimation curves for pure silicate and carbon-rich grains ($\mathcal{P}\simeq 0$) is displayed in Fig.\,\ref{fig:Tsub}, together with that of porous carbonaceous grains mimicking mixtures of Si and C for which Si has sublimated. 
We observe that porosity tends to increase $T\dma{sub}$, as do smaller sublimation timescales. These are interpreted respectively as a loss of efficiency in the sublimation process for lower grain densities, and as a consequence of the exposure time. 
Compared with the usual constant $T\dma{sub}$ approximation, large and small grains have their sublimation temperatures re-evaluated by as much as plus or minus $\sim$20\%.

\subsection{Timescales}\label{sec:timescales}
From what precedes, we are able to define a size- and composition-dependent estimate for the sublimation temperature of a dust grain. Fig.\,\ref{fig:Tsub} reveals an important deviation from the constant sublimation temperature. 
However, we first need to know on which timescale the sublimation has to be considered.
One can notice that an order of magnitude error on the timescale estimate will result only in a modest shift in the sublimation curve. 
Sublimation can be expected to alter the steady-state grain size distribution because it introduces an additional (size-dependent) destruction mechanism; nonetheless we take advantage of the fact that we are using a parametric model and we address only the question:
how big must a grain be to survive a temperature $T\dma{sub}$ for some time $t$.

\begin{figure}[h!]\begin{center}  
\hbox to \textwidth{\includegraphics[angle=0,width=1\columnwidth,origin=bl]{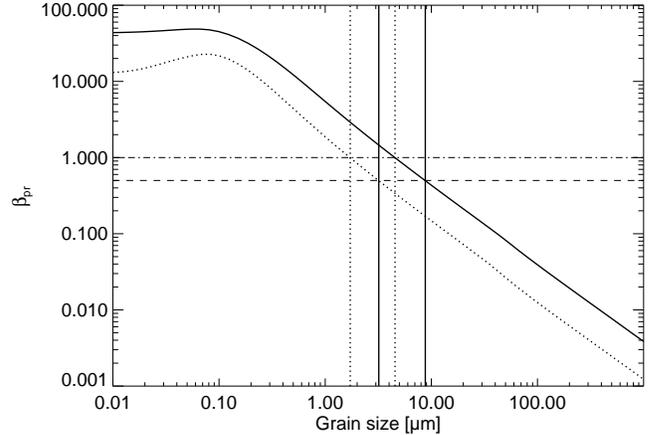}}
\caption{$\beta$ ratio of a dust grain as a function of its size in the Fomalhaut environment. A 50-50 carbon-silicate mixture is assumed and depicted by the lower curve while the upper curve considers half porous carbon grains. The vertical and horizontal lines marks the $\beta$ = 0.5 (blowout size for grains released by parent-bodies on initially circular orbits) and $\beta$ = 1.0 (blowout size for any initial orbit) limits, before and after the sublimation of silicates.}\label{fig:beta_pr}
 \end{center}\end{figure}

A grain lifetime is limited by its removal processes. 
In a debris disk, the main removal process for bound grains is generally destructive collisions, but when the optical depths and/or the dust stellocentric distance is sufficiently small, Poynting-Robertson (PR) drag can become the dominant effect \citep{2005A&A...433.1007W}. Unbound grains are placed on hyperbolic orbits and ejected from the system before they are destroyed: their survival timescale can be equaled to a ``blowout timescale". Assuming near-circular orbits for the parent-bodies, the limit between bound and unbound grains is set by 
$a\dma{blow} = a({\beta = 1/2})$ (Fig.\,\ref{fig:beta_pr}), where $\beta$ is the size-dependent ratio of radiation pressure to gravitational forces.

Here we want to determine the \textit{survival timescale} of a grain, in the \textit{sublimation zone} so depending on its size $a$. 
The \textit{sublimation zone} is defined, for each of the materials that the grains are made of, as the interval $[d_1, d_2]$ between the minimum and maximum sublimation distances (all grain sizes considered) assuming constant sublimation temperature. $d_1$ is essentially independent of the maximum grain size $a\dma{max}$ because the sublimation distance does not vary with size for grains larger than a few $\mu$m (Fig.\,\ref{fig:dTsub}).

We calculate a preliminary grid of models for each grain composition (with no fine computation of the sublimation temperature), and we identify the grain properties and disk surface density that best fit the observations. 
This provides an estimate of the vertical optical depths required to calculate the collisional timescale: the mean timescale a barely bound grain ($\beta \simeq 1/2$) can survive in a collision-dominated disk: 
\begin{equation}
t\dma{col}\uma{0}(a\dma{blow}) = \dfrac{t\dma{orbit}}{2\pi\tau\dma{\perp}\uma{geo}}
\end{equation}
\noindent with $t\dma{orbit}$ the orbital period and $\tau\dma{\perp}\uma{geo}$ the vertical geometrical optical depth at distance $d_1$, a distance representative of the sublimation distance of bound grains.

For larger grains, the collision timescale is scaled with size as \citep{2007A&A...472..169T}
\begin{equation}\label{eq:tcol}
t\dma{col}(a) = t\dma{col}\uma{0} \left(\dfrac{a}{a\dma{blow}}\right)^{0.3}
\end{equation} 
where $t\dma{col}$ depends on the sublimation zone of a given material and on the assumed surface density profile.

We define the PR drag timescale as the time needed for bound grains to spiral from the outer edge of the sublimation zone, to the inner edge of the sublimation zone \citep{2005A&A...433.1007W}
\begin{equation}
 t\dma{PR}(a) = 400\,\beta(a)\frac{{(d_2 - d_1)}^2}{M\dma{*}}.
\end{equation}

As soon as they are produced, unbound grains ($\beta > 1/2$) are placed onto hyperbolic orbits, they are ejected from the sublimation zone and eventually from the field of view. In the sublimation zone, the hyperbolic orbit is approximated by rectilinear uniform motion; we assume the grains are produced at the inner edge of the sublimation zone $d_1$ with initial velocity $v\dma{Kep}(d_1)$. A grain will travel typically a distance between $\sqrt{{d_2\uma{2}}-{d_1\uma{2}}}$ ($\beta(a)=1$) and $d_2 - d_1$ ($\beta \gg 1/2$).
An estimate of the blowout timescale is given by the average between these two extremes (see Eq.\ref{eq:t_blow_0} and Appendix\,\ref{app:t_blow} for a more general estimate)

\begin{equation}
 t\dma{blow} = \dfrac{1}{2} \, \left[\dfrac{{({d_2\uma{2}-d_1\uma{2}})}^{0.5}}{{v\dma{Kep}}} + \dfrac{d_2 - d_1}{v\dma{Kep}}\right].  
\end{equation}

Eventually, the sublimation timescale is equaled to the survival timescale, namely the longest time a grain can be exposed to sublimation, according to

\begin{equation}
t\dma{sub}(a) = \left\{
\begin{array}{rcl}
 &\textrm{min}(t\dma{col}(a),t\dma{PR}(a))&,\textrm{for } a > a\dma{blow}\\
 &t\dma{blow}&,\textrm{for } a \le a\dma{blow}
\end{array}
\right.
\end{equation}

\begin{figure}[h!tbp]\begin{center}  
  \includegraphics[angle=0,width=1.0\columnwidth,origin=bl]{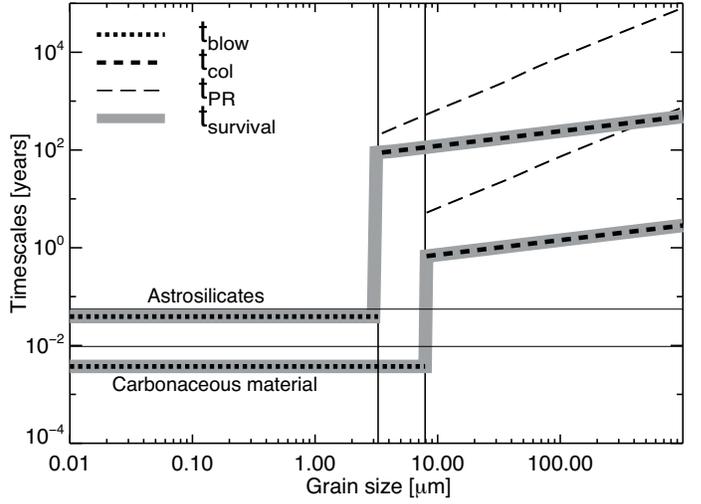}
  \caption{Examples of survival timescales against collisions, PR drag and radiation pressure blowout calculated for an exemplary grain composition ($v\dma{C} = v\dma{Si}$ for silicate, $\mathcal{P}=50\%$ for carbonaceous material). For each effect, the upper curve corresponds to silicate and the lower curve to carbon, each in its sublimation zone and for the disk properties derived from preliminary modeling. The horizontal black lines give the orbital periods for reference. The vertical lines mark the blowout size before (left) and after (right) silicate sublimates.}\label{fig:timescales} 
 \end{center}\end{figure}

A representative example for Fomalhaut is shown in Fig.\,\ref{fig:timescales}. 
The properties of the silicate population and of the carbon population described in details in Sec.\,\ref{sec:results} have been assumed. $t\dma{sub}(\mathrm{C})$ is always smaller than $t\dma{sub}(\mathrm{Si})$ because of the respective locations of the two grain populations. 
The sharp discontinuity between the survival timescale of bound and unbound grains translates into a big jump in the sublimation temperatures and then the distances at the blowout limit.
We stress that Fig.\,\ref{fig:timescales} only illustrates the survival time in the sublimation zone, it should not be used to find the dominant mechanism in the entire dust disk. 

\subsection{Sublimation distances}

The GRaTeR code handles the optical properties of various materials. Broad-band spectra and excesses attributed to warm circumstellar dust disks are well matched using materials of the silicate family or carbonaceous materials. For each of them, we use respectively the thermodynamical quantities associated to amorphous olivine (MgFeSiO$_4$) and graphite (C). They are indicated in Tab.\,\ref{tab:parameters}.

%
\begin{table}[tpb]\caption{Material properties. The thermodynamical constants A and B are tabulated values from \citet{2009A&A...506.1199K} and \citet{1973JChPh..59.2966Z}}\label{tab:parameters}
\begin{tabular}{cccc}
\hline\hline
 & Carbonaceous material &  Silicates &  \\ 
Nickname & C &  Si &  \\ 
 \hline
\multicolumn{3}{l}{\textbf{Thermodynamical properties}} &\\
Material type & Graphite &  Olivine \\ 
\hline
$A$ [cgs]			&  37215  	& 28030  \\
$B$ [cgs] 			& 	1.3028 	& 12.471 	\\
$\mu$\,[$m\dma{p}$] 		& 12.0107	& 172.2331 	\\
$\rho$\,[g.cm$\uma{-3}$] 	& 1.95		& 3.5	 	\\
\hline
\multicolumn{3}{l}{\textbf{Optical properties}} &  \\
Material type & aC ACAR & Astrosilicates\\ 
 Reference & \cite{zubko} & \citet{Draine2003} \\ 
\hline
\end{tabular}
\vspace*{-0.3cm}
\end{table}

\begin{figure}[h!tbp]\begin{center}  
  \includegraphics[angle=0,width=0.8\columnwidth,origin=bl]{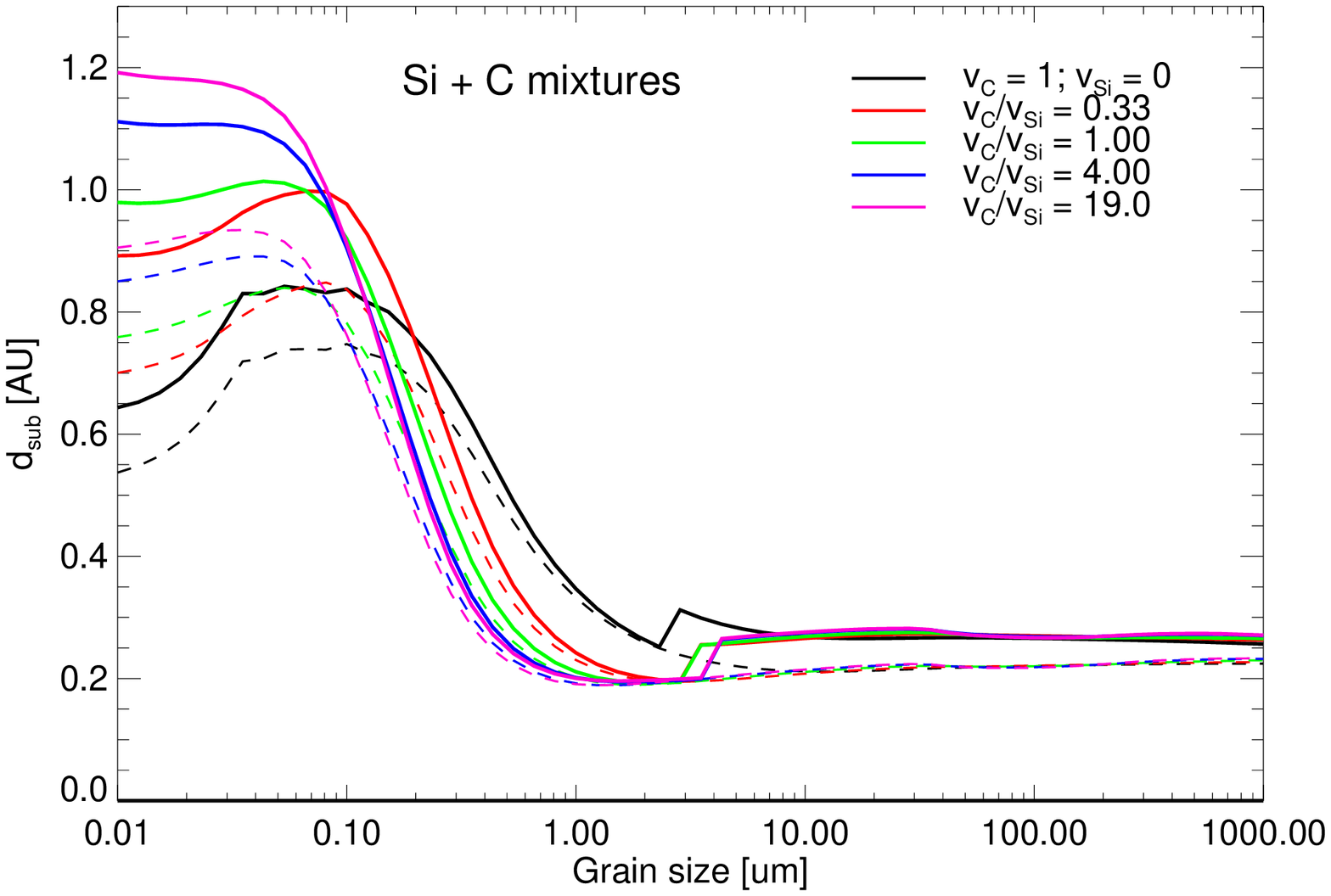}
    \includegraphics[angle=0,width=0.8\columnwidth,origin=bl]{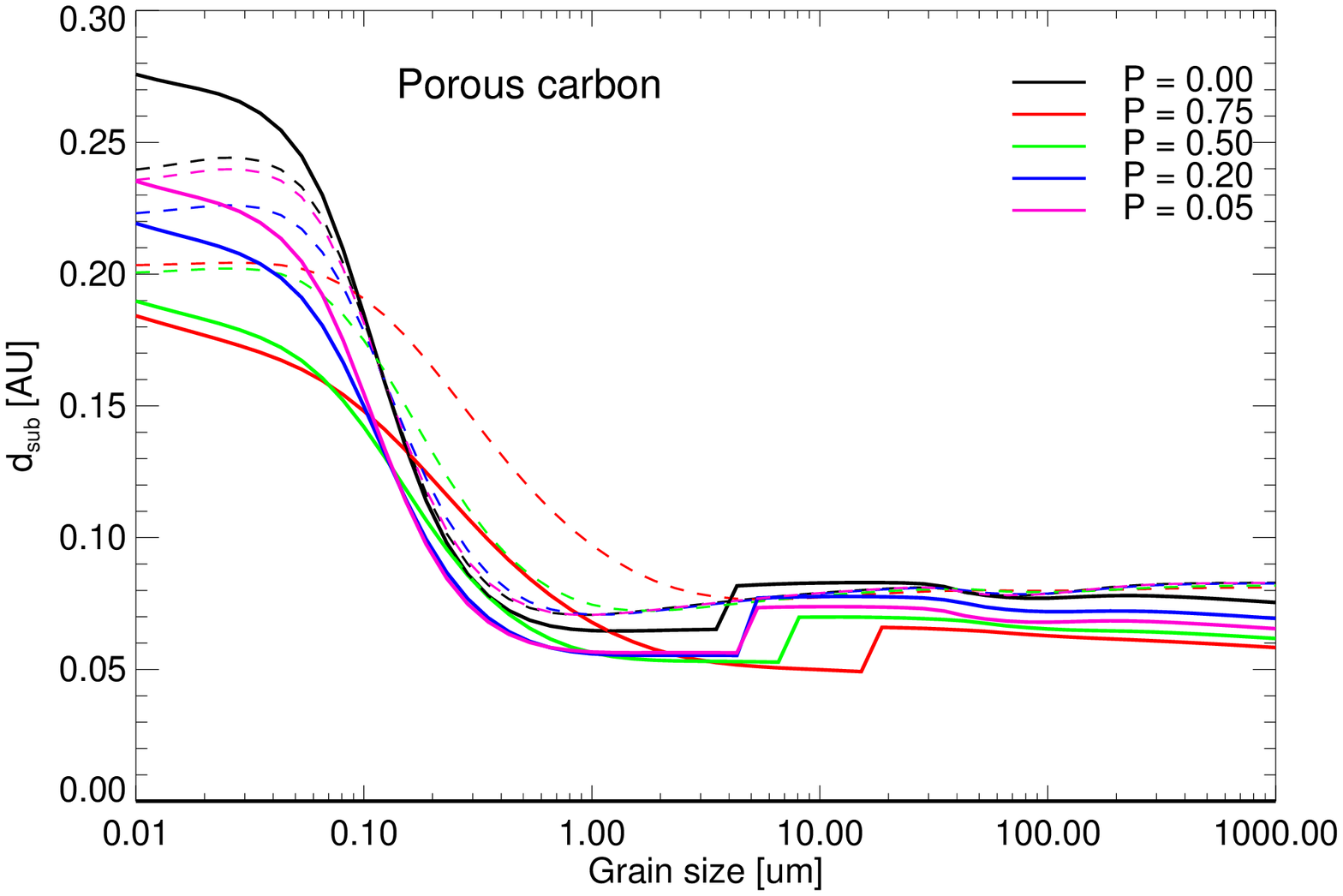}
     \caption{Sublimation distances of silicate (\textbf{top panel}) and carbonaceous material (\textbf{bottom panel}) in composite grains (C+Si or C+vacuum respectively) as a function of grain size in the Fomalhaut environment, with either a constant (dashed line) or a size-dependent (solid line) sublimation temperature. 
Relevant disk properties obtained from preliminary modeling are used (Sec.\,\ref{sec:results}). Size-dependent sublimation temperatures are a function of the size-dependent survival timescales shown in Fig.\,\ref{fig:timescales} causing jumps at the material-dependent blowout limit.}\label{fig:dTsub} 
 \end{center}\end{figure}

After equaling the sublimation timescale to a timescale relevant for each grain (Sec.\,\ref{sec:timescales}), we calculate the thermal equilibrium using the usual formula (Eq.\,\ref{eq:Tdust}). This provides the size dependent-sublimation distance presented in Fig.\,\ref{fig:dTsub} for a few representative material mixtures. The overall shape of the curves is dictated by the thermal equilibrium distances. The net effect of the new model is generally an alteration of the sublimation distances of the small ($< 10 \mu m$) grains, depending on the timescales used for each model. The inner edge of the sublimation is close to 0.2 AU for silicate, 0.05 AU for carbon, while their outer edge lies respectively at 0.7 - 1.2 AU, 0.18 - 0.25 AU respectively.

\subsection{Sublimation model: summary}
We presented an innovative model that is used to calculate size-dependent sublimation distances of dust grains depending on their compositions, porosities and survival timescales. The method can be summarized as follows
\begin{enumerate}
\item A preliminary grid of models is adjusted to the data assuming fixed sublimation temperatures and solving the size-dependent equilibrium temperatures. This provides estimates of the dust disk location and density (Paper II).
\item Size-dependent survival timescales are estimated by searching for the most efficient effect between destructive collisions, PR drag and photo-gravitational blowout (Fig.\,\ref{fig:timescales}). 
\item Size-dependent sublimation temperatures are calculated from the kinetic theory of gases and thermodynamics as a function of the timescale for when a grain is exposed to sublimation (Fig.\,\ref{fig:Tsub}).
\item This sublimation timescale is equaled to the survival timescale for each grain yielding a size-dependent sublimation temperature that accounts for each grain specific dynamical regime. In particular, this causes a discontinuity at the blowout limit (Fig.\,\ref{fig:beta_pr}).
\item Finally, size-dependent sublimation distances are re-evaluated by solving the equilibrium distance of each grain knowing its specific sublimation temperature (Eq.\,\ref{eq:Tdust}).
\end{enumerate}

%

\section{Application and results}\label{sec:results}
\subsection{Observational constraints and star properties}
The model presented above is now confronted with the observations of Fomalhaut already presented in Table\,3 of Paper II. Preliminary analysis of the near-infrared VLTI/VINCI excess, of mid-infrared KIN null excesses, complemented by (low accuracy) spectrophotometric data, suggested the coexistence of two populations of dust in the Fomalhaut environment located within the field of view of the interferometers ($\sim4$ AU and $\sim2$ AU HWHM for FLUOR and KIN respectively).
This architecture is imposed by the high level of the K-band excess with respect to N-band, and the inversion of the null excess slopes upon $\sim10 \mu$m, with a rising profile toward the longer wavelengths.
\citet{2009ApJ...703.1188S} observed a similar break around 10\,$\mu m$ in the Keck null depths of {51\,Ophiuchus}. They were only able to reproduce it with a double-population of dust grains, however with much larger spatial scales due to the distance of the star.
In Paper II, the two dust populations of Fomalhaut were adjusted separately and the model suffered from an inaccurate prescription for the sublimation distances. In the present study we propose a self-consistent modeling of the exozodi and we characterize precisely the properties of the emitting grains and their location. We assume the disk position angle and inclination to equal those of the cold ring ($\mathrm{i} = 65.6\degre$,  $\mathrm{PA} = 156\degre$). A NextGen photosphere \citep{1999ApJ...512..377H} is scaled to the V magnitude of Fomalhaut ($m\dma{V} = 1.2$\,mag) in order to model the total flux received by the grains. It also serves to estimate the excesses attributable to the disk in the photometric data. The interferometric observables on the other hand are independent from the chosen star spectrum.\\

\subsection{Data and modeling strategy}\label{sec:model_st}
\begin{figure}[h!tbp]\begin{center}
  \includegraphics[angle=0,width=1.0\columnwidth,origin=bl]{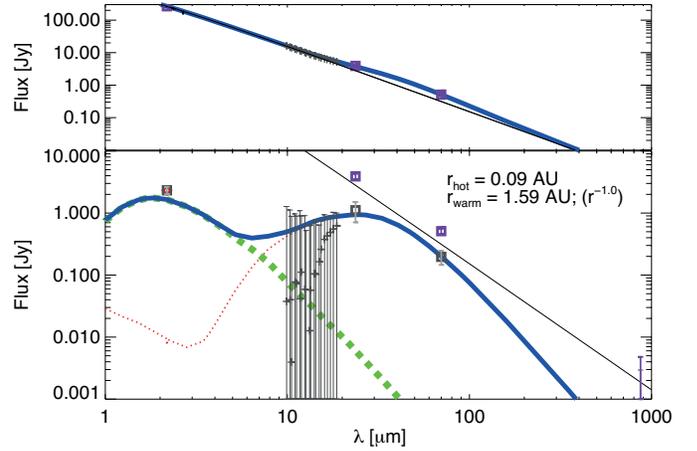}
     \caption{Measured spectral energy distribution, and SED of the best-fitting double-dust belt model (green: hot ring, red: warm belt, blue: total). \textbf{Top panel}: global SED including a NextGen photosphere model (solid black line), \textbf{bottom panel}: circumstellar excess emission. From left to right, thick squares denote the VLTI/VINCI 2.18$\mu$m excess, MIPS 24 $\mu$m and \textit{Herschel}/PACS 70 $\mu$m photometry with 1$\sigma$ error bars. \textit{Spitzer}/IRS spectrum (grey crosses, not fitted) and ALMA 870 $\mu$m photometry (purple cross) are shown as $3\sigma$ upper limits.} \label{fig:sed}\end{center}\end{figure}

\begin{figure}[h!tbp]\begin{center}   
\includegraphics[angle=0,width=0.8\columnwidth,origin=tl]{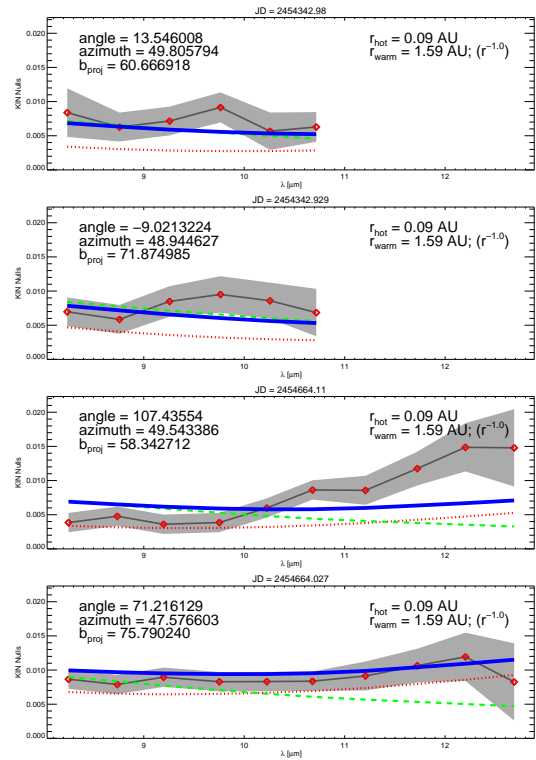}
\caption{Measured KIN excess null depths and nulls of the best-fitting double-dust belt model (green: hot ring, red: warm belt, blue: total) for the four data subsets. The grey regions denote the 1$\sigma$ confidence intervals on the data.}\label{fig:kin} 
 \end{center}\end{figure}

We adopt a strategy in which we fit (1) the ``hot dust ring" ($\lesssim$ 0.4 AU) probed by the shortest wavelengths data, using the same data subset as Paper II (most importantly the VINCI 2.18 $\mu m$ excess), and then (2) the ``warm dust belt" probed by the KIN nulls from 8 to 13 $\mu$m and mid- to far-infrared photometric measurements of the warm ``on-star" excess (Fig.\,\ref{fig:sed} and \ref{fig:kin}). 

These photometric measurements are derived from unresolved observations of the inner Fomalhaut debris disk, well differentiated from the contribution of the cold belt by 
\textit{Spitzer}/MIPS at 23.68$\mu m$ \citep[$F\dma{24} = 3.90\pm{0.40}$\,Jy, ][]{2004ApJS..154..458S}, 
\textit{Herschel}/PACS at 70$\mu m$ \citep[$F\dma{70} = 0.51\pm{0.05}$\,Jy, ][]{2012A&A...540A.125A} 
and ALMA at 870 $\mu$m \citep[taken as a $3\,\sigma$ upper limit: $F\dma{850} < 4.8$\,mJy, ][]{2012ApJ...750L..21B};
 their spatial location is constrained by the instruments point spread functions / beam to be smaller than $\sim 20$AU.
Additional \textit{Spitzer}-IRS spectroscopic data is available:
a small excess emission ($\lesssim 1\,$Jy) shortward of 30 $\mu m$ is reported by \citet{2013ApJ...763..118S}. 
Due to large calibration uncertainties in the absolute photometry ($\sim 10\%$) compared with the interferometric measurements, we do not attempt to fit this spectrum. The IRS spectrophotometry is represented with $3\sigma$ upper limits in Figure\,\ref{fig:sed} to verify the compatibility of the model at these wavelengths. 

The GRaTeR code is used to adjust the dataset in several steps. We point out that all the results presented below correspond to thermal emission by the dust and that the scattering of the stellar spectra by the grains is always negligible (although systematically calculated). \\

As a first step, we assess the hot population, assumed to be composed of compact carbonaceous grains, because they are very small and lie within the silicate sublimation zone (Paper II).
We adopt the same grid of models as the one presented in Table\,4 of Paper II, yielding $\sim$200,000 models for 5 free parameters: the minimum grain size $a\dma{min}$, the slope of the size distribution $\kappa$, the surface density peak $r\dma{0}$, outer slope $\alpha\dma{out}$ and the total disk mass $M\dma{dust}$ (with the maximum grain size fixed to 1 mm). The inner density slope is assumed to be steep ($\alpha\dma{in}=+10$) - but this has no impact as this region is within the sublimation radii of carbons as we will see. 

As a second step we focus on the disk's warm component and take advantage of the linearity of the data that allows us to sum the contributions from the two components. Using the same parameter space, we adjust simultaneously 6 free parameters: the mass, geometry and disk properties of the warm component and the mass of the hot component (modeled in step 1) to the 36 measurements composing the full dataset (VINCI, KIN, MIPS, PACS, ALMA). The warm ring is made of a 50-50 mixture of astronomical silicates and carbonaceous material characteristic of the material properties commonly inferred from debris disks and solar system asteroids studies. We do not vary this parameter as it is found to be essentially unconstrained by the observations in the previous study.

Two approaches are used. 
In the first approach, the outer slope of the density profile is taken as a free parameter, and the inner slope is assumed to be very steep (fixed to $+10$). 
In the second approach, the outer slope is fixed to $-1.5$ - close to the best-fit for the first approach and consistent with the profile expected for a collisional equilibrium under the effect of size-dependent radiation pressure
 \citep{2008A&A...481..713T} - and we vary the inner slope $\alpha\dma{in}$. \\

A $\chi^2$ minimization is performed as well as a Bayesian statistical analysis to measure the likelihood of the model parameters.
The best-fitting parameters and Bayesian estimates discussed in this section are summarized in Tab.\,\ref{tab:morphology} and the resulting models are shown in Fig.\,\ref{fig:sed} and \ref{fig:kin}. Probability curves are shown in Appendix B. \\

 \subsection{The hot population}

As expected, the hot component is mostly constrained by the near-infrared data and best explained by a very narrow ring of $\gtrsim 10$ nm refractive grains, with a density profile peak matching the sublimation distance of carbons ($r\dma{0} = 0.09$ AU), where only the largest grains survive. 
The best model ($\chi^2_{r}$ = 1.4 with 20 degrees of freedom, \textit{d.o.f.}) is found for a minimum grain size matching the smallest values of the parameter space ($a\dma{min}=10$nm, $\kappa = -6$), although the Bayesian analysis favors $a\dma{min} = 20$ nm. 
In fact, the exact properties of this ring have little impact on the resulting SED, as long as the grain size-dependent temperatures and total mass are consistent with the K-band excess. 
The slopes of the density profile and of the size distribution are qualitatively very steep but their exact values are not well identified due to the strong dependence of temperature with distance and grain size. 
However, the data also carries \textit{spatial} information. In particular the dust must be confined within the  (Gaussian) field-of-view of the VLTI \textit{and} it must not let too much emission through the complex KIN transmission map in the mid-infrared. 
Grains larger than $\sim {\lambda}/{2\pi} \simeq 0.3 \mu$m are also inefficient emitters in the K-band, and due to the single power law used, very small grains can become dominant.
For these reasons and despite the modeling degeneracy between grain size and disk location, the minimum grain size cannot exceed a few $\sim$10 nm. For instance, the best model found with $a\dma{min} = 1 \mu$m has a $\chi^2\dma{r}$ of 2.9. This result is reminiscent of what was found by \citet{2011A&A...534A...5D} who show that the exozodi of Vega must be composed of grain much smaller than $1\,\mu$m based on multiwavelengths constraints in the near-IR.
We use $a\dma{min} = 10$ nm as a working assumption for the rest of the study although what really matters is to determine which of the grains are the emitters.

In the inner solar system, the dust probed by the NASA's deep Impact mission likely consists of $\sim20\mu$m-sized highly porous dust aggregates \citep[\textit{e.g.}][]{2013A&A...550A..72K}.
The hot exozodi of Fomalhaut rather consists of nanometer to submicrometer grains that we interpret as the elemental monomers produced after the break-up of larger dust particles. As demonstrated in Paper II (Fig.\,6 and Tab.\,5), models with higher porosity provide less good fit to the data. For instance, setting the porosity to 85\,\% yields a smallest $\chi^2\dma{r}$ of 1.8 for $a\dma{min} = 1.5\mu$m. A large amount of sensitive measurements would be needed to go beyond this solid carbon grain model \citep[see \textit{e.g.}][]{Lebreton2012a}.

\begin{figure}[h!b]\begin{center}  
  \includegraphics[angle=0,width=1.0\columnwidth,origin=bl]{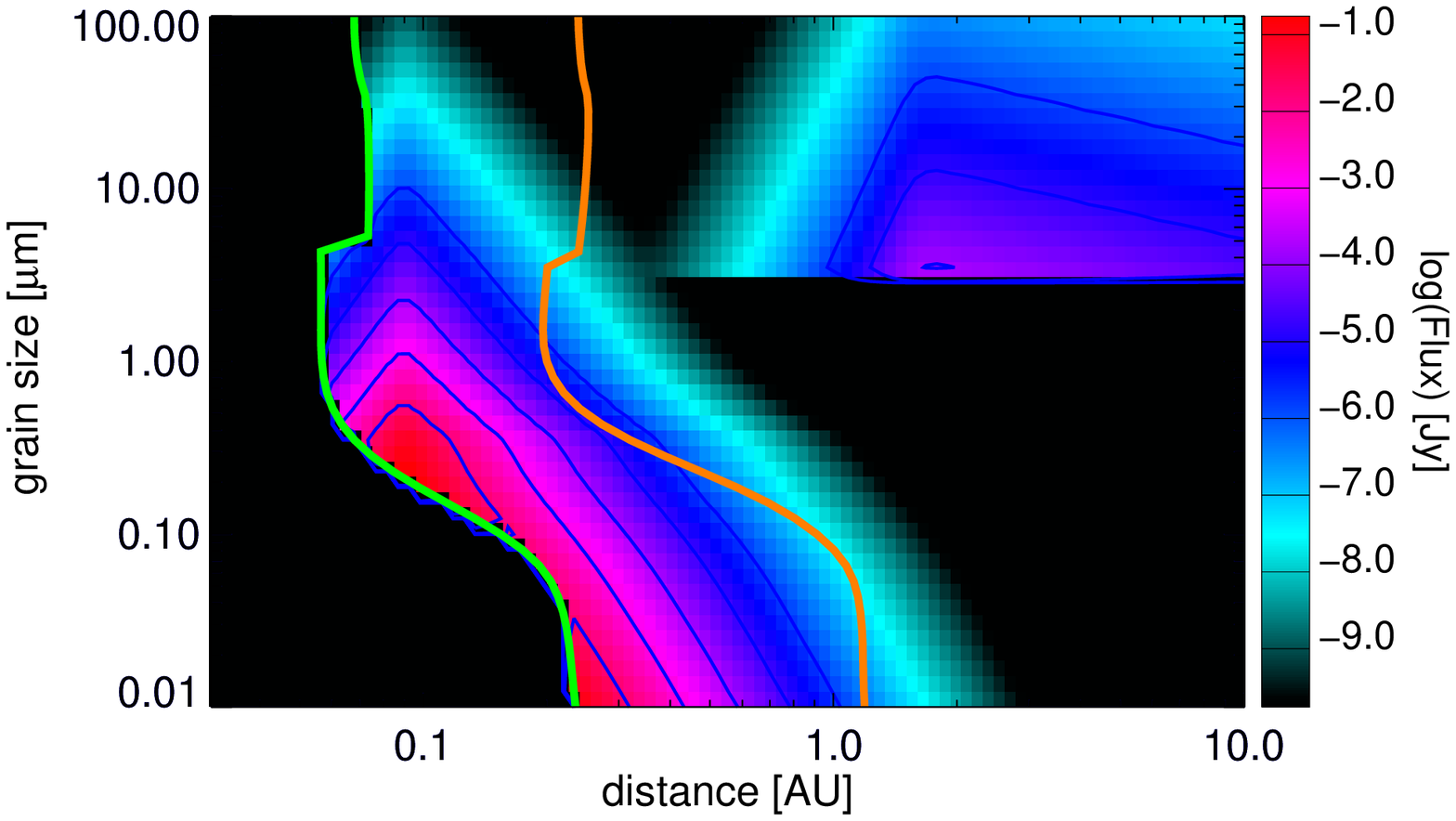}
  
\vspace{-1.5cm}  
  
    \includegraphics[angle=0,width=1.0\columnwidth,origin=bl]{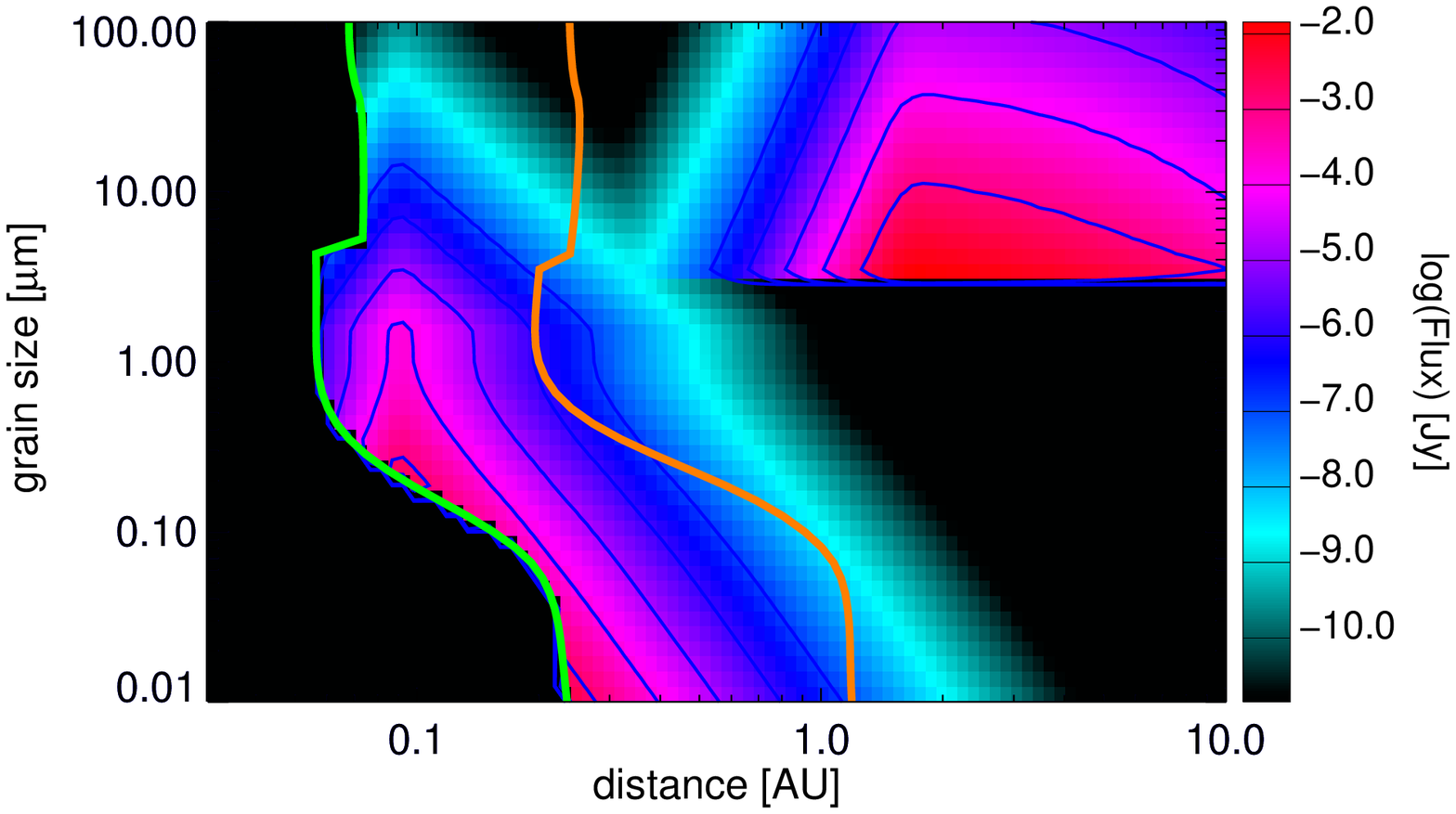}
\vspace{-1.cm}  
     \caption{Maps showing the absolute distribution of flux as a function of distance from the star and grain size. \textbf{Top panel}: $\lambda$ = 2.18 $\mu$m, \textbf{bottom panel}: $\lambda$ = 12.0 $\mu$m. The orange and green lines mark the sublimation distances of silicate and carbon grains respectively. }\label{fig:map} 
 \end{center}\end{figure}

The new sublimation prescription provides a more reliable estimate of the dust location with respect to the constant sublimation temperature assumption (Fig.\,\ref{fig:map}). 
With $\alpha\dma{out} = -6$ (\textit{i.e.} a very narrow ring), the sublimation distance of 0.01 $\mu$m grains is 0.235 AU ($\simeq d_2$); for these grains the emission falls down to 10\% of the peak flux at 0.34 AU. 
Grains larger than $a\dma{min}$, in particular those in the range 0.1-0.5 $\mu m$ that lie close to $d_1$ contribute to thermal emission in similar proportions compared with 10 nm grains. 
Thus, independent from our choices for the parameter space limits, the emission is by far dominated by unbound grains and a robust result is that this hot exozodi is composed of grains smaller than $\sim$0.5 $\mu$m. 
The dust mass is dominated by the smallest grains, due to $\kappa < -4$. 
We would like to stress that forcing the material sublimation temperatures to higher values would not yield better fit to the data as it would only move the peak emission toward even shorter wavelengths (Tab.\,\ref{tab:morphology}, Fig.\,\ref{fig:sed}).
In the following, we adopt the above parameters for the hot ring and keep the total dust mass as the single free parameter for this component (Tab.\,\ref{tab:morphology}).

\subsection{The warm population}
 
With the first approach (variable outer slope, fixed inner slope), the analysis indicates that the mid- to far-infrared data is best fitted by a dust belt peaking at $r\dma{0}$ = 1.6 AU and declining slowly as $r^{-1}$. The minimum size is close to the blowout limit ($a\dma{blow}=8.8 \mu$m), in a rather steep distribution ($\kappa < -3.8$), yielding a reduced $\chi^2$ of 1.56 (with 30 \textit{d.o.f.}). These properties are consistent with the theoretical expectations that the warm population is produced through a collisional cascade in a parent-body belt, although the steep distribution might be more indicative of a recent catastrophic collision than a steady-state debris disk. 

Interestingly, the second approach (fixed outer slope, variable inner slope) yields similar results to those of the first one, except that a second family of solutions arises, with very small grains located at the outer edge of the range of explored peak radii. Nonetheless this solution can be excluded based on the absence of silicate features in the \textit{Spitzer}/IRS spectrum that would be created by such tiny silicate grains. We add prior information in the Bayesian analysis to reject solutions with $a\dma{min} < a\dma{blow}/10$ (Appendix B.2) and find a best reduced $\chi^2$ of 1.60. Due to the different geometrical profile assumed, the peak radius is found to lie further out, at $\sim 2.5$ AU. The inner ring is not as steep as previously assumed ($\alpha\dma{in} = +3$), which is suggestive of an inward transport of material by Poynting-Robertson drag mitigated by destructive collisions \citep[See for example ][]{2012A&A...537A.110L}. 

\subsection{modeling summary}
\begin{figure}[h!]\begin{center}  
\includegraphics[angle=0,width=0.9\columnwidth,origin=tl]{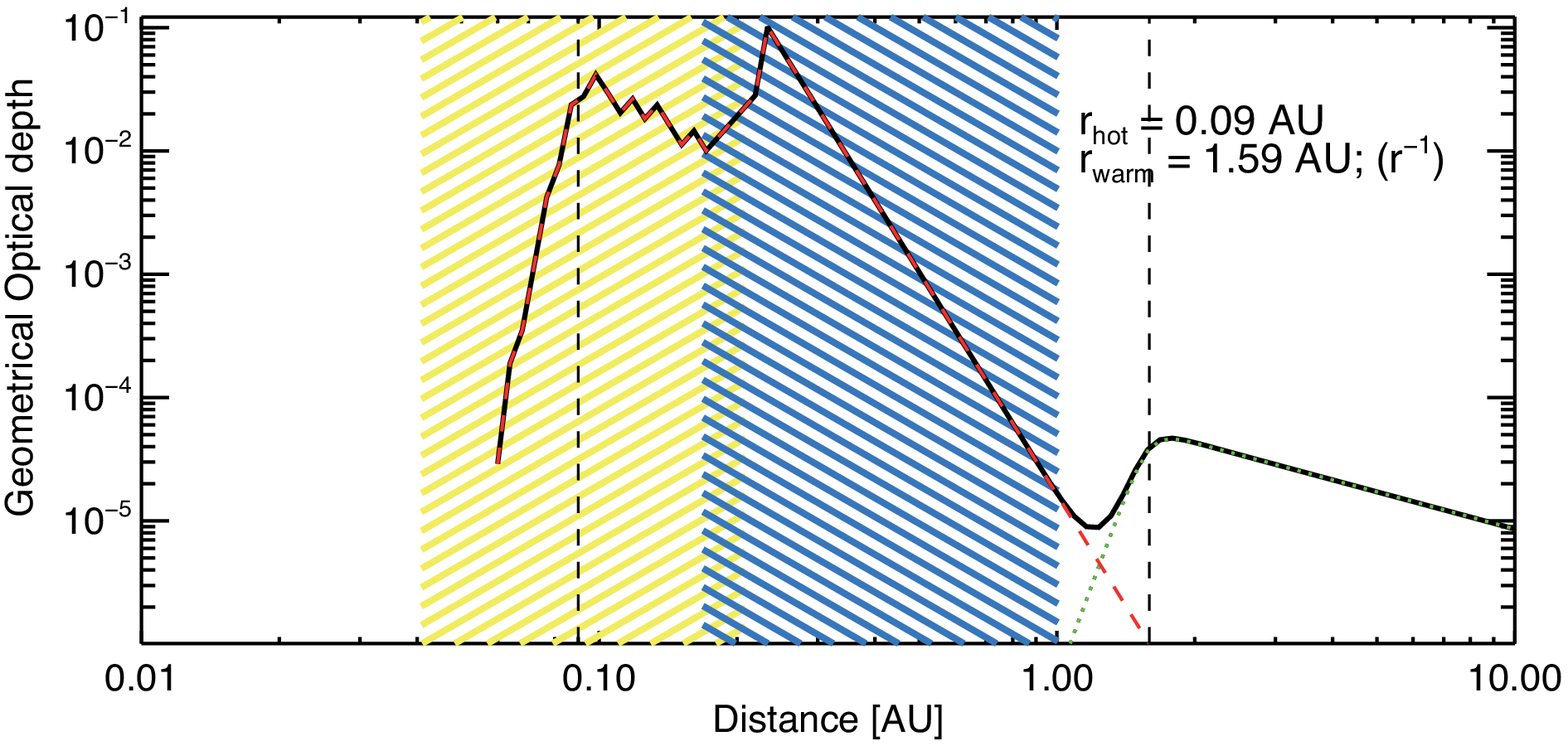}

\vspace{-0.3cm}

\includegraphics[angle=0,width=0.9\columnwidth,origin=tr]{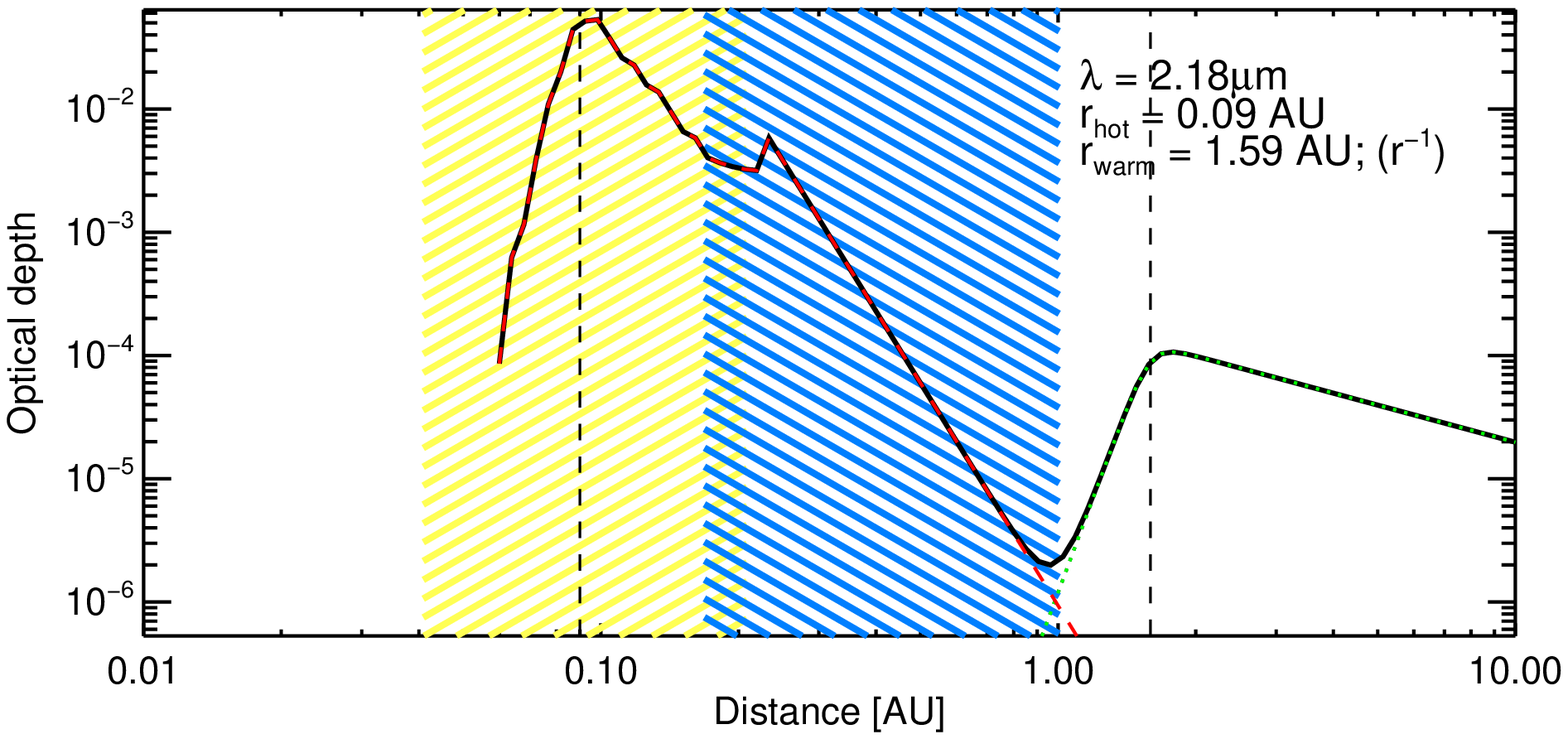}

\vspace{-0.3cm}

\includegraphics[angle=0,width=0.9\columnwidth,origin=bl]{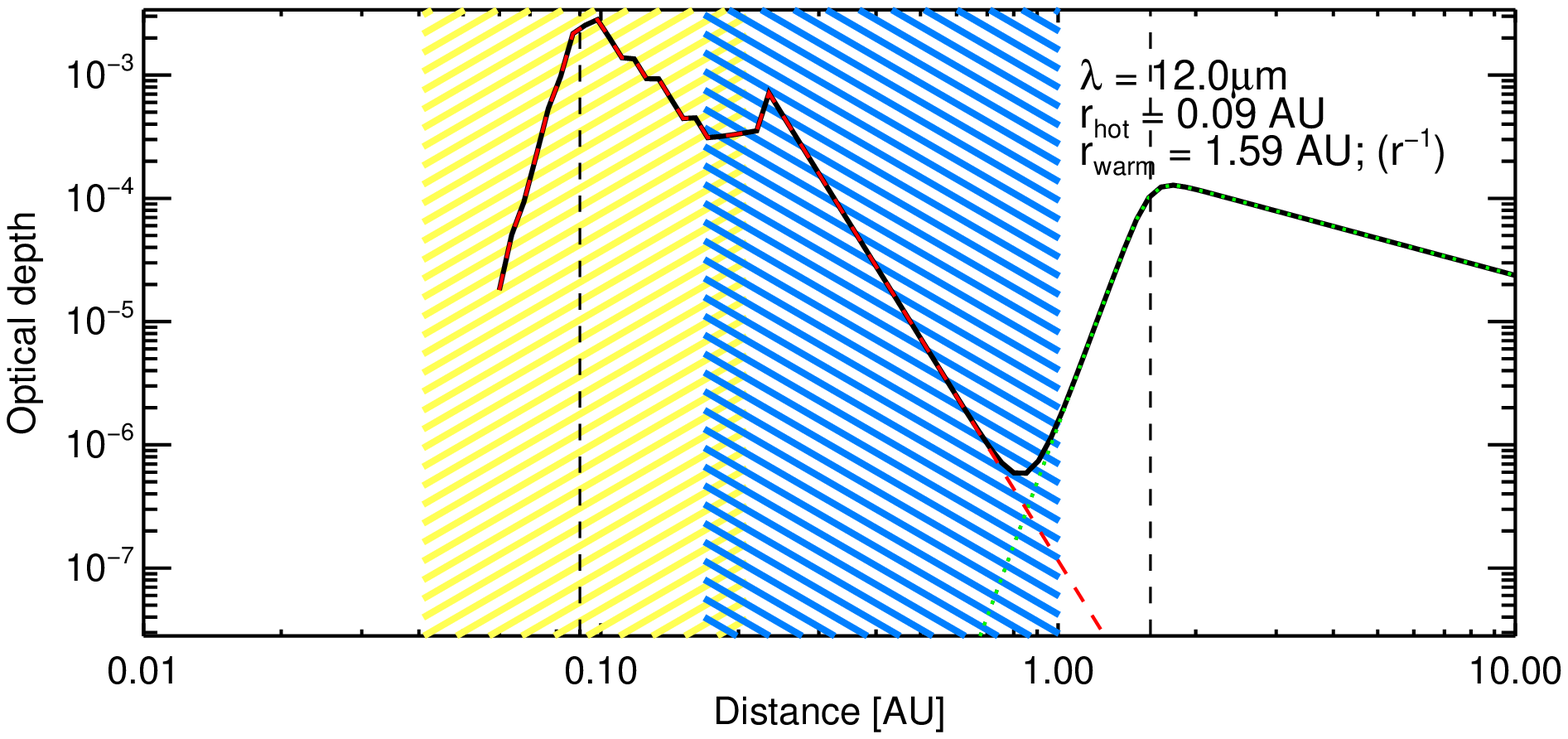}
\vspace{-0.3cm}
\caption{Radial profiles for the best-fitting double-ring model (approach 1. Red: hot ring; green: warm belt): geometrical vertical depth profiles (\textbf{top panel}), optical depth profiles at $\lambda=2.18 \mu$m (\textbf{middle panel}) and $\lambda=12\mu$m (\textbf{bottom panel}). The dashed regions represent the sublimation zones of silicates (blue region) and carbons (yellow). The effect of size-dependent sublimation is evidenced by the ragged profile of the hot ring optical depth. A logarithmic sampling is used with 56 sizes ranging from 0.01 to 1000 $\mu$m.}\label{fig:tau_prof} 
 \end{center}\end{figure}

Fig.\,\ref{fig:sed} shows an excellent fit to the spectral energy distribution. The near-infrared excess is almost solely produced by the hot ring at the sublimation distance. The warm belt at 2 AU is sufficient to explain the measured SED from mid- to far-infrared wavelengths and it is compatible with the \textit{Spitzer}/IRS excesses.
Furthermore the null excesses are mostly produced by the hot component below $\sim\,10\,\mu$m in the 2007 configuration, but their rising profiles at longer wavelengths requires the contribution from the warm component. All the model excesses are compatible with the measurements within $\sim 1\sigma$ except for the short equivalent baseline in the 2008 sample that is underpredicted by $\sim 2\sigma$ upon 11$\mu$m. A possible interpretation would be the existence of an azimuthal asymmetry in the disk.
Overall the warm ring is much more massive than the hot ring with respective values of $\sim\,2 \times\,10^{-6}\,M\dma{\oplus}$ and $\sim\,2.5 \times\,10^{-10}\,M\dma{\oplus}$ in grains smaller than 1 mm, but the hot ring is significantly brighter (Fig.\,\ref{fig:map}) due not only to its high temperature, but also to the nature of its constituent grains. The tiny grains that the hot ring is made of constitute a much larger effective cross-section than an equivalent mass of material concentrated in large grains. 
The resulting profiles are illustrated in Figure\,\ref{fig:tau_prof} that shows the geometrical vertical optical depth and effective vertical optical depths at two representative wavelengths. 

In summary, the hot ring at a fraction of AU consists essentially in grains that seem to contradict the dynamical/collisional theoretical constraints ($a \ll a\dma{blow}$, $\kappa \le -5.0$, steep density profile). 
The ``warm belt" at about 2 AU seems compatible with the ``classical" debris disk picture ($a\dma{min}$, $\kappa$ and $\alpha\dma{out}$ are compatible with blowout of small grains and a collisional size distribution).

%
\begin{table*}[h!tpb]\begin{center}\caption{Properties of the Fomalhaut exozodiacal disk derived from the fit to the data. \textit{Approach 1} labels the model with free outer slope, \textit{approach 2} that with free inner slope. Numbers in bold are the parameters of the smallest $\chi^2$ fits, while the intervals give the $1\sigma$ confidence intervals from the Bayesian analysis. The * exponent marks the fixed parameters.}\label{tab:morphology}
\begin{tabular}{cccccc}
\hline\hline
\vspace{0.1cm}\\
Properties & \multicolumn{2}{c}{Hot ring} & \multicolumn{2}{c}{Warm belt} \\ 
~ & \textit{Approach 1} & \textit{Approach 2} & \textit{Approach 1} & \textit{Approach 2}\\ 
\hline
\vspace{0.1cm}\\
\multirow{2}*{Density peak $r_0$ [AU]}  & 0.09$\uma{*}$ (0.23$\uma{[1]}$) & 0.09$\uma{*}$ (0.23$\uma{[1]}$) & \textbf{1.59} & \textbf{2.52} \\
										&  & ~ & [1.62, 2.43] & [1.54, 2.15]  \\
\vspace{0.1cm}\\
Inner density slope $\alpha\dma{in}$  & $+10\uma{*}$ & $+10\uma{*}$ & $+10\uma{*}$ & \textbf{+3} \\
									  & 			 & 				 & 				& [+2.99, +7.52] \\
\vspace{0.1cm}\\
\multirow{2}*{Outer density slope $\alpha\dma{out}$}  & $-6\uma{*}$ & $-6\uma{*}$ & \textbf{-1.0} & $-1.5\uma{*}$ \\
			   										  & &  & [-3.62, -0.60] & \\
\hline
\vspace{0.1cm}\\
\multirow{2}*{Mass up to 1 mm [$M\dma{\oplus}$]}& \textbf{2.5\,.\,10$\uma{\textbf{-10}}$}	& \textbf{2.6\,.\,10$\uma{\textbf{-10}}$} & \textbf{2.86\,.\,10$\uma{\textbf{-6}}$} & \textbf{1.87\,.\,10$\uma{\textbf{-6}}$} \\ 
												 &  &  & $[6.12\,.\,10^{-7}, 9.53\,.\,10^{-5}]$ & $[1.14\,.\,10^{-6}, 7.81\,.\,10^{-5}]$ \\
Maximum surface density [cm$\uma{-2}$]		& {7.24\,.\,10}$\uma{{11}}$ & {7.46\,.\,10}$\uma{{11}}$ 	& 110	&	 66.6 \\
Maximum V-band optical depth $\tau\dma{\perp}$ $\uma{[3]}$		& \textbf{2.2\,.\,10}$\uma{\textbf{-3}}$  & \textbf{2.3\,.\,10}$\uma{\textbf{-3}}$ 	& \textbf{9.8\,.\,10}$\uma{\textbf{-5}}$	&	 \textbf{1.2\,.\,10}$\uma{\textbf{-4}}$ \\
Fractional luminosity $L\dma{D}/L\dma{\star}$	& $5.89\,.\,10^{-3}$ & $5.71\,.\,10^{-3}$ & $5.04\,.\,10^{-4}$&  $6.09\,.\,10^{-4}$ \\
\hline
\vspace{0.1cm}\\
Composition &  C	& C &	Si+C & Si+C \\
\vspace{0.1cm}
\multirow{2}*{Minimum grain size [$\mu$m]}	&	0.01	& 0.01	& \textbf{3.51}  & \textbf{2.31} \\
											&	 	&  	& [1.37, 56.1] & [0.99, 57.2] \\
\multirow{2}*{Grain size slope}	&	-6	& -6	& \textbf{-4.8} & \textbf{-4.06} \\
								&	    &   	& [-6.0, -3.8] & [-6.0, -3.7] \\
Typical temperatures [K]$\uma{[2]}$& [2000, 2200] & [2000, 2200] & [403,464] & [326,364] \\
\hline
$\chi\uma{2}$ (dof = 30)  & --	& -- & 46.9  & 47.9 \\
\hline\vspace*{-0.3cm}
\end{tabular}\end{center}
\vspace*{-0.3cm}
{\sc Notes -- } 
$\uma{[1]}$ Approximate density peak position of the \textit{smallest} grains (they vanish below that distance). \\
$\uma{[2]}$ Give range of temperatures at $r_0$ from the smallest to 1 mm grains. For the hot ring, the temperature of the smallest grains is given at their sublimation distance. 
$\uma{[3]}$ The optical depth is directly proportional to the dust mass and therefore constitute another definition to this free parameter (given in V-band by convention).
\end{table*}

 \vspace{-0.3cm}

%% file: rik5_sec5.tex
\section{Origin of the dust}\label{sec:origin}

In this section, we test several mechanisms as possible explanations for
the peculiar morphology suggested by the modeling results.
First, we discuss the viability of parent belts at the
locations of the hot ring and the warm belt (Sec.\,\ref{sec:parent}).
Subsequently, we review possible replenishing mechanisms for the warm belt (Sec.\,\ref{sec:replen_warm}),
and test whether the pile-up of dust in the sublimation zone can explain the hot ring (Sec.\,\ref{sec:accumulation}).
In Sec.\,\ref{sec:release},
we examine whether the observed population of unbound grains
can be produced by the disruption of larger bodies due to sublimation.
Finally, we investigate the role of gas in retaining the small grains in the hot ring
by slowing down both their blowout and sublimation (Sec.\,\ref{sec:gas}).
A short summary of our findings is presented in Sec.\,\ref{sec:origin_sum}.

To compare theory with observation, we
approximate the fractional luminosity of dust at radial distance $r$
as the fraction of the star covered by dust at that location:
\begin{equation}
  \label{eq:fraclum_crosssection}
  \frac{ L\dma{D} }{ L\dma{\star} } (r)
  \approx \frac{ \sigma\dma{D}(r) }{ 4 \pi r^2 }.
\end{equation}
Here, $ \sigma\dma{D} (r) $ is the collective cross section of the dust at radial distance $r$.
When the excess flux
can be assumed to be
entirely due to
spherical grains of a single size $a$
(and hence with a mass of $ m = 4 \pi \rho\dma{d} a^3 / 3 $),
the fractional luminosity
is related to the total dust mass according to
\begin{equation}
  \label{eq:fraclum_mass}
  \frac{ L\dma{D} }{ L\dma{\star} } (r)
  \approx \frac{ 3 M\dma{D} }{ 16 \pi \rho\dma{d} a r^2 }.
\end{equation}

\subsection{In-situ dust production through a collisional cascade?}\label{sec:parent}

Excess infrared emission from debris disks is normally interpreted as
thermal emission from dust produced
through mutual collisions between larger bodies in a planetesimal belt.
Therefore, we first test whether the observed NIR excess can be explained by
the production of small dust grains by asteroid belts at the locations of
the hot ring and the warm belt.

In steady-state collisional evolution, a planetesimal belt
at a given radial distance from the star
can only
contain
a maximum amount of mass at any given time,
because more massive belts process their material faster \citep{2003ApJ...598..626D, 2007ApJ...658..569W}.
Assuming that the size distribution follows the classical \citet{1969JGR....74.2531D} power law
($ \kappa = -3.5 $, valid if the critical specific energy for dispersal $ Q_D^{\star} $ is independent of particle size) at all sizes, and extends down to the blowout size,
the mass corresponds to
a maximum fractional luminosity
of \citep{2007ApJ...658..569W}
\begin{multline}
  \label{eq:max_fraclum1}
  \max \left[ \frac{ L\dma{D} }{ L\dma{\star} } \right]
  = 7.0 \times 10^{-9}
  \biggl( \frac{ r }{ \mathrm{1~AU} } \biggr)^{7/3} \\
  \quad \mbox{} \times
  \biggl( \frac{ dr/r }{ 0.5 } \biggr)
  \biggl( \frac{ a\dma{c} }{ \mathrm{30~km} } \biggr)^{0.5}
  \biggl( \frac{ Q_D^{\star} }{ \mathrm{150~J~kg}^{-1} } \biggr)^{5/6}
  \biggl( \frac{ e }{ 0.05 } \biggr)^{-5/3} \\
  \mbox{} \times
  \biggl( \frac{ M\dma{\star} }{ \mathrm{1.92~M}\dma{\sun} } \biggr)^{-5/6}
  \biggl( \frac{ L\dma{\star} }{ \mathrm{16.63~L}\dma{\sun} } \biggr)^{-5/6}
  \biggl( \frac{ t\dma{age} }{ \mathrm{440~Myr} } \biggr)^{-1}.
\end{multline}
Here, we inserted fiducial values for
the relative width of the planetesimal belt $ dr/r $, the radius of the largest bodies $ a\dma{c} $,
the
critical specific energy for dispersal
$ Q_D^{\star} $, and the mean planetesimal eccentricity $ e $.\footnote{
These values were found to give a good fit to debris disks around a sample of A stars \citep{2007ApJ...663..365W},
and can be used as first order estimates for these poorly constrained parameters in the case of exozodiacal dust.}
For the stellar mass $ M\dma{\star} $, the stellar luminosity $ L\dma{\star} $,
and the age of the system $ t\dma{age} $,
we used the parameters of Fomalhaut found by \citet{2012ApJ...754L..20M}.
Cratering collisions, which have a specific energy lower than $ Q_D^{\star} $,
lead to an increased erosion of large bodies, and therefore a faster processing of the available material,
and a lower maximum fractional luminosity at any given age.
However, they are not accounted for in the model of \citet{2007ApJ...658..569W}.
Including cratering collisions would lower the numerical factor in Eq.~\ref{eq:max_fraclum1}
by about a factor 4 to 5 \citep{2010Icar..206..735K}.
Our choice of $ Q_D^{\star} = \mathrm{150~J~kg}^{-1} $ can be seen as a conservative estimate.

Evaluating Eq.~\ref{eq:max_fraclum1} at
the radial locations of the hot ring ($ r \approx 0.25$~AU) and the warm belt ($ r \approx 2$~AU),
gives maximum fractional luminosities of
$ 2.7 \times 10^{-10} $ and $ 3.5 \times 10^{-8} $, respectively.
We note that the modeling results indicate steeper size distributions ($ \kappa < -4.0 $) for both components,
which would yield much higher maximum fractional luminosities \citep[using Eq. 16 of][]{2007ApJ...658..569W}.
However, the observations only probe the lower end of the size distribution,
and it is unlikely that the steep power law extends all the way to parent body sizes.
In contrast, the size distribution is expected to be shallower at large sizes,
where the strengthening of bodies due to self-gravity becomes important.
Furthermore, the theoretical $ \kappa = -3.5 $ is confirmed observationally
for km-sized objects in the solar system's asteroid belt \citep{1969JGR....74.2531D}.
In a more thorough discussion of the stringentness of the maximum fractional luminosity,
\citet{2007ApJ...658..569W} find that by pushing the parameters,
the fractional luminosity can be made to exceed the maximum given by Eq.~\ref{eq:max_fraclum1} by a factor 100 at most.
Since the fractional luminosities derived from modeling the interferometric data (Tab.~\ref{tab:morphology})
are more than two orders of magnitude higher than the maximum ones,
we conclude that neither of the two components can be explained
by in-situ asteroid belts that have been in collisional equilibrium for the age of the system.

\subsection{Replenishing the warm belt}\label{sec:replen_warm}

If the warm belt is in steady state (and not a transient phenomenon),
some mechanism must operate to replenish the observed dust,
other than an in-situ collisional cascade.
A possible source of the material is the outer cold belt at about 140~AU.
We now proceed to
estimate the rate at which the dust needs to be replenished,
and then examine whether PR drag from the cold belt is capable of providing this mass flux.

The warm belt is found to have a relatively steep size distribution ($ \kappa < -4.0 $),
and therefore its total dust mass is dominated by the smallest grains present.
In the warm belt these are particles close to the blowout size,
which are destroyed by collisions on a timescale of
several thousands of years (Eq.~\ref{eq:tcol}).
With a dust mass of a few times $ 10^{-6} \; \mathrm{M}\dma{\earth} $,
the mass flux through the warm belt must be of the order of
$ 10^{-9} \; \mathrm{M}\dma{\earth} \; \mathrm{yr}^{-1} $.

PR drag can only supply a limited amount of material,
since grains undergo mutual collisions as they migrate inwards,
and the fragments produced in these collisions can be blown out.
This was demonstrated by \citet{2005A&A...433.1007W},
using a model that assumes a single particle size, fully destructive collisions, and circular orbits.\footnote{
The model of \citet{2005A&A...433.1007W} ignores stellar wind drag,
but this process is not expected to be important for Fomalhaut and other A stars,
because their mass loss rates are predicted to be very low.}
Based on this model, we can estimate the collision-limited PR drag mass flux
from a dust source located at $r\dma{source}$ inward to a radial distance $r$,
which is
\begin{equation}
  \label{eq:max_mass_flux}
  \max \left[ \dot{ M }\dma{PR}(r) \right]
  = \frac{ L\dma{\star} \sqrt{ G M\dma{\star} } \beta Q\dma{pr} }
    { 2 c^3 \left( \sqrt{ r\dma{source} } - \sqrt{ r } \right) }.
\end{equation}
Here, $L\dma{\star}$ is the stellar luminosity, $G$ is the gravitational constant, $M\dma{\star}$ is the stellar mass,
$Q\dma{pr}$ is the wavelength-averaged radiation pressure coefficient, $c$ is the speed of light.
A detailed derivation of this equation will be presented by R. van Lieshout et al. (in prep.).
This maximum is independent of the amount of material at the dust source.
Also, its dependence on grain properties is through $ Q\dma{pr} $ and $\beta$,
for which we know $ 0 < Q\dma{pr} < 2 $,
and $ \beta < 0.5 $ for dust released from large ($ \beta \approx 0 $) parent bodies on circular orbits.

We evaluate Eq.~\ref{eq:max_mass_flux} at the location of the warm belt ($ r \approx 2$~AU),
using $r\dma{source} = 140$~AU, the stellar parameters of Fomalhaut,
$ Q\dma{pr} = 1.0 $, and $\beta$ = 0.5.
The $\beta$ ratio is that of the smallest bound grains,
which are dragged in the most efficiently, and can therefore provide the highest mass flux.
The resulting maximum mass flux
is about $ 1.2 \times 10^{-12} \; \textrm{M}\dma{\oplus} \; \textrm{yr}^{-1}$,
which is several orders of magnitude smaller than the mass flux required to maintain the warm belt.
This indicates that PR drag is not the main replenishing mechanism for the warm belt.\\

\vspace*{-0.4cm}
\begin{figure}[!htbp]
\begin{center}%
  \includegraphics[angle=0,width=0.9\columnwidth,origin=bl]{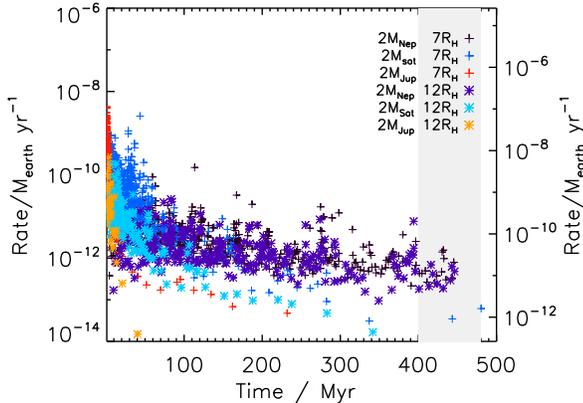}
\vspace*{-0.4cm}
     \caption{Steady-state time evolution of the rate at which material is scattered inward of 2 AU through planet scattering, in Earth masses per year; left axis: $M\dma{outer} = 4 M\dma{\oplus}$, right axis: $M\dma{outer} = 105 M\dma{\oplus}$. Several compact planetary configurations are tested with with 7 or 12 equal mass (2 Jupiter- 2 Neptune- or 2 Saturne-mass planets), equally spaced by 7 or 12 Hill-radii. The grey area represents the estimated age of Fomalhaut.}\label{fig:amy_planetscat} 
\end{center}
 \end{figure}

Another possible mechanism for transporting material from the outer cold belt to the warm belt
is the inward scattering of small bodies by planets \citep{2012MNRAS.420.2990B,2012arXiv1209.6033B}.
The outer belt mass is constrained to be between 4 $M\dma{\oplus}$ \citep{2012ApJ...750L..21B} and 110 $M\dma{\oplus}$ \citep{2012A&A...540A.125A} and the star age is 440$\pm{40}$Myr \citep{2012ApJ...754L..20M}.
If a chain of planets were to orbit between the cold and warm belts, these planets could scatter small bodies inwards, in essence resupplying the warm and hot belt with material.
\citet{2012arXiv1209.6033B} used N-body simulations to investigate this scattering process and determined a maximum flux of scattered particles as a function of time. This maximum occurs if a chain of tightly packed, low mass planets, were to orbit between the belts. Here, we apply these models to the case of Fomalhaut, scaling the simulations to account for the stellar mass of $2.1M_\odot$ and the inner edge of the outer belt at 133AU. The results are shown in Fig.\,\ref{fig:amy_planetscat}. 
The important information given by this plot is the maximum possible rate at which material can be scattered inwards, given by the upper envelope of the scattered points. By assuming that the material is efficiently converted to small dust, this also gives us the maximum rate at which the observed small dust in the hot and warm belts could be resupplied. 
This process could potentially provide a mass flux
of up to $5\times 10^{-11} \; \mathrm{M}\dma{\oplus} \; \mathrm{yr}^{-1} $
continuously for the age of the system, which is still not sufficient to compensate the collision rate ($4\times 10^{-10} \; \mathrm{M}\dma{\oplus} \; \mathrm{yr}^{-1} $). 
This value is very much a maximum mass flux, as it was calculated using the planet configuration that is the most efficient at scattering material inwards and it assumes that the scattered objects are entirely disrupted upon entering the inner regions of the Fomalhaut planetary system.

\subsection{Accumulation of sublimating dust grains?}\label{sec:accumulation}

For the hot ring,
the radial distribution of dust exhibits a strong peak in surface density in the carbon sublimation zone,
with much lower levels of dust further out (see Fig.~\ref{fig:tau_prof}).
This spatial profile suggests the existence of a mechanism to confine
the carbonaceous dust in the sublimation zone.

\citet{2009Icar..201..395K} predicted that the dual effect of PR
drag and radiation pressure blowout can result in
an accumulation of grains around the sublimation radius.
In the following we briefly explain the pile-up mechanism
in terms of the three stages identified by \citet{2009Icar..201..395K}.
(1) Initially, the grains are far away from the star, and sublimation is negligible.
The grains spiral inward due to PR drag and gradually heat up as they come closer to the star.
(2) As the dust temperature approaches the sublimation temperature,
the grains start to lose mass due to sublimation.
As a consequence of the increasing cross-section-to-mass ratio of the dust grains,
radiation pressure gains in relative importance compared to gravity
(i.e. the $ \beta $ ratio of the dust grains becomes higher).
This increases the eccentricity of the dust orbits,
and therefore their orbital size.
Hence, the inward radial migration is slowed down.
Eventually, the decrease in semi-major axis due to PR drag is compensated by
the increase due to sublimation.
This happens roughly at the radial distance
where the PR drag timescale equals the sublimation timescale \citep{2008Icar..195..871K}.
(3) Finally, the size of the dust grains drops below the blowout radius,
and their orbits become unbound.
At this point the grains either leave the system,
or they fully sublimate before they exit the sublimation zone.
The net outcome of this process is an accumulation of dust in the
sublimation zone, and this result is very robust against various grain
properties (composition, porosity, fractal structure).

The pile-up mechanism is a result of the interplay between PR drag and sublimation,
and was investigated for drag dominated disks (i.e. disks in which collisions are unimportant).
If collisions
are significant,
they may inhibit the process of dust pile-up.
Furthermore, a significant pile-up requires very low orbital eccentricities $ e \lesssim 10^{-2}$ \citep{2008Icar..195..871K, 2011EP&S...63.1067K}.
PR drag can circularize the orbits of dust particles,
but only if the source region is distant enough.
These caveats will be examined in further detail by R. van Lieshout et al. (in prep.).

The material in the pile-up needs to be replenished from further out by PR drag.
Assuming that a dust source is located at the radius of the warm component ($r\dma{source} =$ 2~AU),
without specifying
how this source can be maintained,
we now use Eq.~\ref{eq:max_mass_flux} to
estimate how much
material can be transported inward to the hot ring at $r =$ 0.23~AU.\footnote{
In the best fit model, two populations of dust contribute equally to the hot component emission,
one at $ r = 0.1 $~AU, and one at $ r = 0.23 $~AU.
The latter is most optimistic choice for the PR drag scenario, resulting in the highest mass flux.}
Inserting the parameters of Fomalhaut, together
with $Q\dma{pr} = 1.0 $ and $\beta$ = 0.5, yields a maximum mass flux
of $ 1.4 \times 10^{-11} \; \textrm{M}\dma{\oplus} \; \textrm{yr}^{-1}$.

If the orbital eccentricities of the dust grains are sufficiently low,
a pile-up of dust will occur
at roughly the radial distance
where the PR drag timescale equals the sublimation timescale.
The dust stays in the pile-up until it is completely sublimated,
or sublimated to below the blowout size,
which approximately takes a sublimation timescale.\footnote{
Individual dust grains survive for longer than a sublimation timescale,
since they end up on eccentric orbits that take them out of the sublimation zone.
However, the time they spend in the pile-up, where they are observed,
corresponds to the sublimation timescale.}
Therefore, the dust stays in the pile-up for a PR drag timescale.

A rough estimate of the maximum total mass of piled up dust can be found by
multiplying the maximum mass flux rate found earlier with the PR drag timescale at the pile-up location.
For $r = 0.23$~AU, the PR drag timescale is 22~yr,
resulting in a maximum dust mass of $ 3.1 \times 10^{-10} \; \textrm{M}\dma{\oplus} $,
which is compatible with the modeled hot component mass.
However, the maximum fractional luminosity due to this amount of mass in bound grains at this location (Eq.~\ref{eq:fraclum_mass}),
is only $ 9.2 \times 10^{-6} $, which is several orders of magnitude lower than the value of the best fit model.
This estimate is independent of the amount of material at the dust source,
and only depends on grain properties through $\beta$,
which should not be higher than $ \beta = 0.5 $.
The reason for the discrepancy is that the dust grains found by the model are much smaller
than the $ \beta = 0.5 $ particles considered for the pile-up mechanism,
and therefore constitute much more cross section for the same amount of mass.
Since these small grains are below the blowout size,
they are removed from the system on timescales much shorter than the PR drag timescale.
Hence, the pile-up of dust alone cannot explain the observed excess emission of the hot component.

\subsection{The release of small dust grains in the sublimation zone}\label{sec:release}

The observation of dust grains with sizes far below the blowout size presents a problem.\footnote{
Very small grains sometimes have $ \beta $ ratios below unity,
due to their low optical efficiencies.
However, for the material types tested here, and Fomalhaut as the host star,
$ \beta $ stays well above unity, and the smallest grains are unbound (see Fig.~\ref{fig:beta_pr}).
}
These particles have very short survival timescales,
and therefore have to be replenished quickly.
Their detection in the sublimation zone indicates that they could be released
from unseen larger bodies that fall apart as they sublimate.
The increase in the number of particles, with conservation of total mass,
would lead to an increase in collective cross section,
and the steep dependence of sublimation on temperature
could explain the sharp peak in emission in the sublimation zone.

Larger bodies could be transported into the sublimation zone by various processes,
such as P-R drag, or inward scattering by planets.
Alternatively, the small particles could be released by an evaporating planet
that is present in the sublimation zone for an extended period
and gradually loses material.
For now, we ignore what is the exact mechanism that provides the material,
but rather focus on the mass source term $ \dot{ M } $
required to explain the observed fractional luminosity.

The fractional luminosity of the hot ring can be approximated by Eq.~\ref{eq:fraclum_mass}.
This is possible because its size distribution is very steep,
so the cross section is dominated by the smallest grains,
and because its radial distribution is a sharp spike,
so all grains are roughly located at the same radius.
Furthermore, because of the steep size distribution ($ \kappa < -4.0 $),
the total dust mass in the hot ring is dominated by the smallest grains.
Hence, the modeling results provide a rough constraint on
the product of the mass source term $ \dot{ M } $
and the survival timescale of the small grains $ t\dma{surv} $,
according to
\begin{equation}
  \label{eq:fraclum}
  \frac{ L\dma{D} }{ L\dma{\star} } = \frac{ 3 \dot{ M } t\dma{surv} }{ 16 \pi \rho\dma{d} a r^2 }.
\end{equation}

A closer inspection of the modeling results reveals that
the flux of the hot ring is mainly due to two populations of dust,
which have similar contributions
(see Fig.~\ref{fig:map}).
The reason for this complication is that the radial distribution is truncated at the size dependent sublimation radius.
One population consists of very small grains ($ a = 0.01 \, \mu $m) located at $ r = 0.23 \, $AU,
the other of slightly larger grain ($ a = 0.2 \, \mu $m) at $ r = 0.10 \, $AU.
These slightly larger grains are still well below the blowout size.
In the following we will test both sets of parameters when evaluating the survival timescale. 
For the material density we assume that of carbon with a porosity of 5\%
($ \rho\dma{d} = \mathrm{ 1.85~g~cm^{-3} } $).

We now make an estimation of the survival timescale of the relevant dust grains,
in order to find the required mass production or influx to explain the observed fractional luminosity.
The mechanisms that are critical to the survival of small grains close to a star are sublimation and blowout.
Sublimation destroys the grains on a timescale of
\begin{equation}
  \label{eq:t_subl}
  t\dma{subl} = \frac{ a }{ | \dot{ a } | },
\end{equation}
where $ \dot{ a } $ is given by Eq.~\ref{eq:asub}.
This is the time it takes for a grain to sublimate, assuming the sublimation rate remains constant.
$ t\dma{subl} $ is highly dependent on $ r $ (through dust temperature), and therefore uncertain.
For the blowout timescale, we take
\begin{equation}
  \label{eq:t_blow_0}
  t\dma{ dyn }
  = \sqrt{ \frac{ 5 r^3 }{ 2 G M\dma{\star} ( \beta  - 1 ) } },
\end{equation}
which is the time it takes a particle to move outward from its release point to twice that radial distance,
valid for $ \beta \gg 1 $ (a derivation is presented in App.~\ref{app:t_blow_0}).

\begin{figure*}[h!tbp]\begin{center}   \hspace*{-0.4cm}
  \includegraphics[angle=0,width=2.\columnwidth,origin=bl]{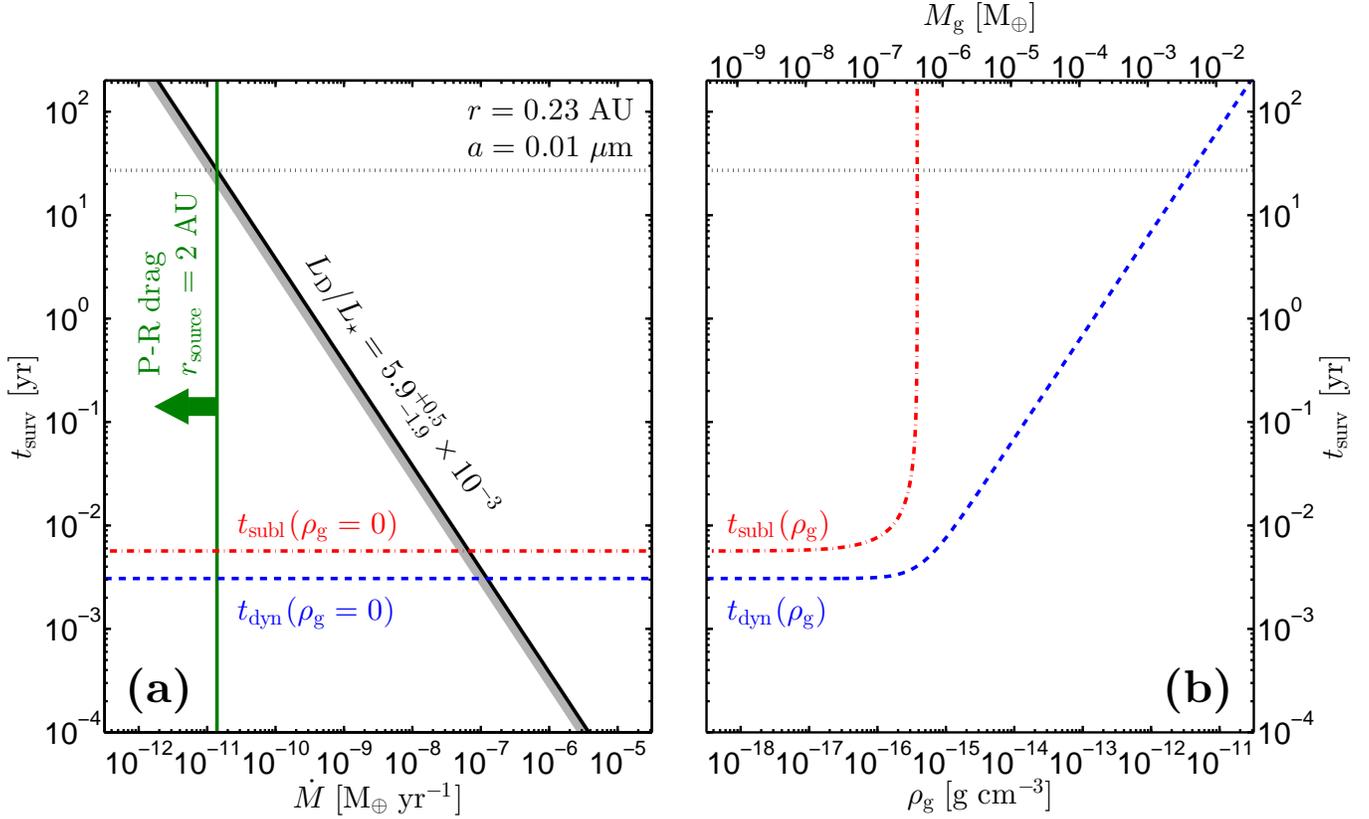}
     \caption{
A summary of the constraints on the hot ring,
assuming it consists of 0.01~$\mu$m carbon grains located at 0.23~AU.
Panel \textbf{(a)}:
The diagonal line shows the relation between the mass flux and the dust grain survival timescale,
as constrained by the observed fractional luminosity (black, with error margins in grey, Eq.~\ref{eq:fraclum}).
The horizontal lines indicate the typical timescales (in a gas free environment) for
destruction by sublimation (red, dash-dotted, Eq.~\ref{eq:t_subl}), and
removal by blowout (blue, dashed, Eq.~\ref{eq:t_blow_0}).
The vertical line with the arrow indicates the maximum mass flux due to P-R drag
from a very dense source region located at 2~AU (green, Eq.~\ref{eq:max_mass_flux}).
Panel \textbf{(b)}:
The dependence on gas density of
the sublimation timescale (red, dash-dotted, Eq.~\ref{eq:t_subl}), and
the blowout timescale (blue, dashed, the sum in quadrature of Eq.~\ref{eq:t_blow_0} and Eq.~\ref{eq:t_blow_rho})
The top axis gives the total gas mass corresponding to the midplane gas densities on the bottom axis,
assuming the gas is located in a vertically isothermal ring of width $ \Delta r = r $ (Eq.~\ref{eq:m_gas}).
The horizontal, black, dotted line
marks the minimum survival time required if the observed material is to be provided by PR drag.
It extends across both panels to show the gas density and total gas mass this would imply.
}\label{fig:tmdot_s001} 
 \end{center}\end{figure*}

\begin{figure*}[h!tbp]\begin{center}   \hspace*{-0.4cm}
  \includegraphics[angle=0,width=2.\columnwidth,origin=bl]{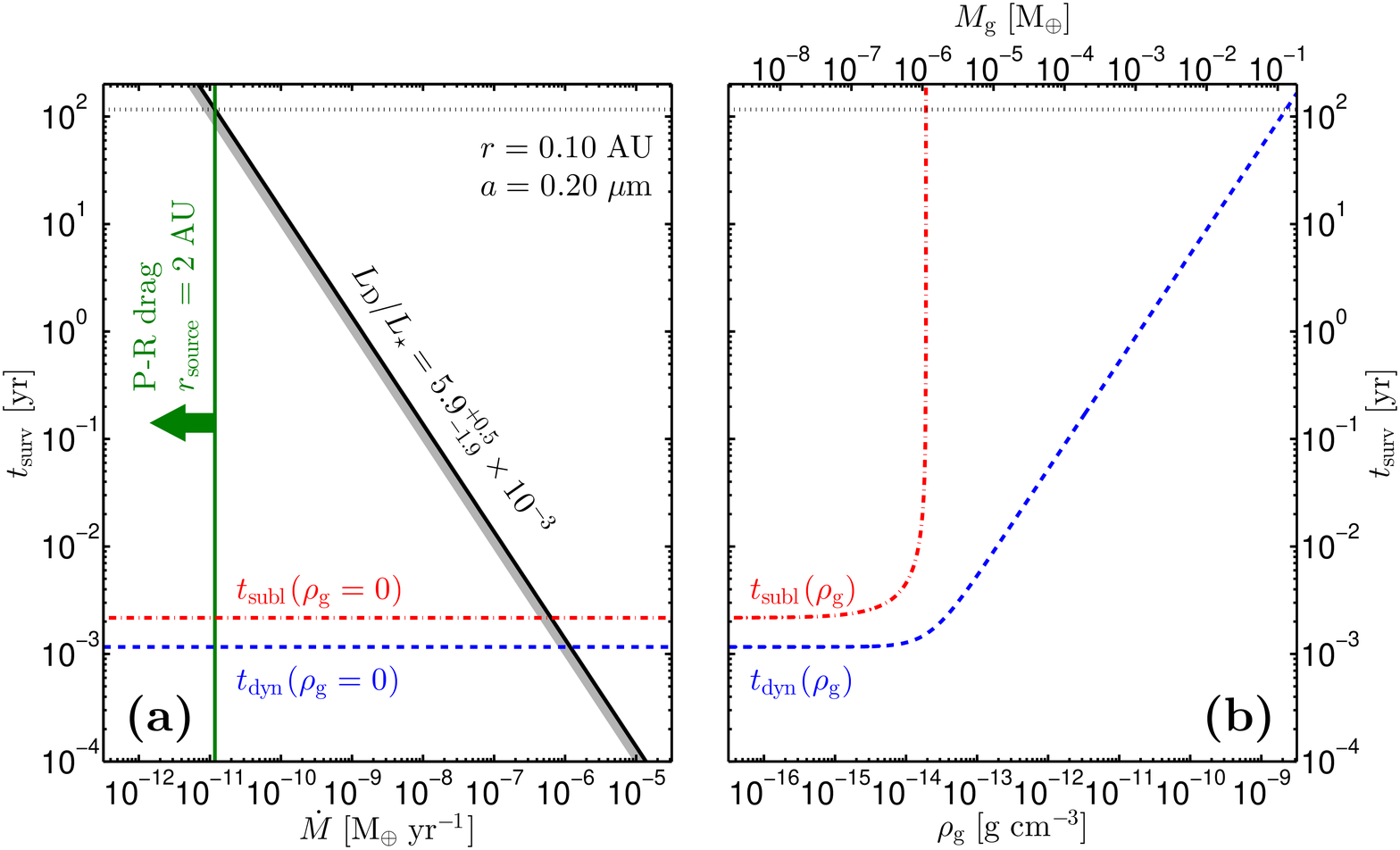}
     \caption{
Same as Fig.~\ref{fig:tmdot_s001}, but assuming the hot ring consists of 0.2~$\mu$m carbon grains at 0.1~AU.
Different scales are used on the horizontal axes in panel \textbf{(b)}.
}\label{fig:tmdot_s02} 
 \end{center}\end{figure*}

Figures~\ref{fig:tmdot_s001}a and \ref{fig:tmdot_s02}a
show the constraints that the observations put on
the product of $ \dot{ M } $ and $ t\dma{surv} $ for
the $ a = 0.01 \, \mu $m grains at $ r = 0.23 \, $AU
and the $ a = 0.2 \, \mu $m grains at $ r = 0.10 \, $AU, respectively.
Also shown are the typical survival timescales of these dust grains,
and the maximum mass influx due to PR drag from a source region at 2~AU.
The figures reveal that a mass source term of at least $ \sim 10^{-7} \; \mathrm{M}\dma{\earth} \; \mathrm{yr}^{-1}$
is required to explain the observations, if the grains survive for $ t\dma{dyn} $.
Comparing this with the maximum mass influx due to PR drag
indicates that this mechanism cannot provide enough material.

%

Other processes that release small dust particles in the sublimation zone
may yield higher mass source terms.
For instance, \citet{2012ApJ...752....1R} report a mass loss rate
from the possible evaporating planet KIC~12557548~b
of $ \sim 10^{-9} \; \mathrm{M_\oplus \; yr^{-1} } $,
sustainable for $ \sim 0.2$~Gyr.
This rate would still be insufficient to explain Fomalhaut's hot ring,
and the dust morphology inferred for KIC~12557548~b is very different from a ring.
However, the actual mass loss rate of an evaporating planet is highly dependent on
variables such as its mass, size, temperature, and composition.

Alternatively, the estimates of the dust survival timescale could be too low.
If the small dust grains would somehow be retained in the sublimation zone,
their survival timescale could be larger.
One possibility is that the assumption that the disk is completely gas free
is not valid in the sublimation zone.
In the next subsection, we explore the influence of gas on the dust survival timescales.

\subsection{The influence of gas on dust survival timescales}\label{sec:gas}

Because dust sublimation converts solid dust to gas,
some gas is expected to be present in the sublimation zone.
The presence of gas increases the blowout timescale,
because particles on their way out are slowed down by gas drag.
Assuming that the subsonic Epstein drag law can be used for the gas drag force on the small dust grains,
we find that the blowout timescale at high gas densities is given by
\begin{equation}
  \label{eq:t_blow_rho}
  t\dma{dyn} ( \Delta r , \rho\dma{g} )
  = \frac{ \rho\dma{g} v\dma{th} \Delta r \;\! r^2 }{ G M\dma{\star} ( \beta - 1 ) \rho\dma{d} s }.
\end{equation}
Here, $ \rho\dma{g} $ is the mass density of the gas,
$ v\dma{th} $ is the mean thermal speed of the gas,
and $ \Delta r $ is the distance to be travelled by the blowout grain
(this equation is derived in App.~\ref{app:t_blow_rho}).
This timescale is shown for $ \Delta r = r $ in Figs.~\ref{fig:tmdot_s001}b and \ref{fig:tmdot_s02}b,
using the gas densities on the lower axes.

By assuming that the gas is confined to a ring around the star
with a radial width of $ \Delta r $, and that it is vertically isothermal,
we can express $t\dma{dyn}$ in terms of the total gas mass,
and eliminate the dependence on $ \Delta r $ and $ v\dma{th} $.
The gas surface density of a vertically isothermal disk with midplane density $ \rho\dma{g} $ is given by
\begin{equation}
  \label{eq:gas_surf_dens}
  \Sigma\dma{g}
  = \frac{ \pi \sqrt{ \gamma } }{ 2 }
    \rho\dma{g} v\dma{th} \sqrt{ \frac{ r^3 }{ G M\dma{\star} } },
\end{equation}
where $ \gamma $ is the adiabatic index,
for which we assume $ \gamma = 1.5 $.
Then, the total gas mass is given by
\begin{align}
  \label{eq:m_gas}
  M\dma{g}
  & = 2 \pi r \times \Delta r \times \Sigma\dma{g} \\
  & = \pi^2 \sqrt{ \gamma } \rho\dma{g} v\dma{th} \Delta r \sqrt{ \frac{ r^5 }{ G M\dma{\star} } },
\end{align}
and the blowout timescale becomes
\begin{equation}
  \label{eq:t_dyn_m_gas}
  t\dma{dyn} ( M\dma{g} )
  = \frac{ M\dma{g} }{ \pi^2 \sqrt{ \gamma G M\dma{\star} r } ( \beta - 1 ) \rho\dma{d} s }.
\end{equation}
This is shown in Figs.~\ref{fig:tmdot_s001}b and \ref{fig:tmdot_s02}b with the gas masses on the upper axes.

In the optimistic case that the hot ring is due to $ a = 0.01 \, \mu $m grains at $ r = 0.23 \, $AU,
the hypothesis that the dust is supplied by PR drag from the warm belt
requires a survival timescale of about 30~yr (the dotted black line in Fig.~\ref{fig:tmdot_s001}).
This timescale is reached at a gas density of approximately $ 4 \times 10^{-12} \; \mathrm{g \; cm^{-3} } $,
which corresponds to a total gas mass of about $ 5 \times 10^{-3} \; \mathrm{M\dma{\earth}} $.
Assuming that all this gas is provided by sublimating dust at the maximum PR drag rate,
this mechanism needs to have operated for approximately
$ M\dma{g} \; / \; \max [ \dot{ M }\dma{PR} ]
\approx 5 \times 10^{-3} \; \mathrm{M\dma{\earth}} \; / \; 1.4 \times 10^{-11} \; \textrm{M}\dma{\oplus} \; \textrm{yr}^{-1}
\approx 0.4$~Gyr, which equals the age of the system, to explain the total gas mass.

The gas will also affect the sublimation of dust grains,
since the presence of gas raises the sublimation temperature
(see Sec.~\ref{sec:sublimation}).
Figs.~\ref{fig:tmdot_s001}b and \ref{fig:tmdot_s02}b also show
the dependence of the sublimation timescale on gas density.
This assumes that the ambient gas only consists of carbon,
so the gas densities on the lower axes correspond to the partial densities used
to compute the sublimation rate in Sec.~\ref{sec:sublimation}.
For high gas densities, the dust particles will not sublimate, but grow.
This would stabilize the dust grains,
possibly explaining their very high temperatures.

\subsection{Summary of theoretical findings}\label{sec:origin_sum}

The distribution of dust in the inner few AU of the Fomalhaut system
that was found by modeling the observations,
is difficult to reconcile with a steady-state model.
Both the hot and warm components contain
more dust than can be explained by
in-situ production via a collisional cascade that has operated for the age of the system.
If these two components are in steady state (as opposed to being transient phenomena),
other processes must be operating to maintain the observed dust populations.
We considered several mechanisms as potential sources of the dust.
Table~\ref{tab:mdot} gives an overview of the mass fluxes that can be attained by these mechanisms.

%
\begin{table}[h!tpb]\begin{center}\caption{Mass fluxes of several potential supply or production mechanisms, and of the main destruction mechanisms (destructive collisions and radiative transfer blowout).}\label{tab:mdot}
\begin{tabular}{cc}
\hline
 Mechanism & Mass flux ($ \mathrm{M}\dma{\earth} \; \mathrm{yr}^{-1} $) \\ 
\hline
\multicolumn{2}{c}{Warm belt} \\
\hline
PR drag from 140~AU & $ < 1.2 \times 10^{-12} $ \\
Planetesimal scattering & $ \lesssim 5\times 10^{-11} $ \\
Collisions 	& $\sim -4\times 10\uma{-10}$\\
\hline
\multicolumn{2}{c}{Hot ring} \\
\hline
PR drag from 2~AU & $ < 1.4 \times 10^{-11} $ \\
Evaporating planet$^{[1]}$ & $ \sim 10^{-9} $ \\
Blowout 	& ~$\sim -8\times 10\uma{-8}$\\
\hline
\end{tabular}\end{center}
\vspace*{-0.3cm}
{\sc Notes -- } $^{[1]}$ This is the value found by \citet{2012ApJ...752....1R} for KIC~12557548~b.
It is given here for comparison.
\end{table}

The warm belt is dominated by barely bound grains that are primarily destroyed by mutual collisions.
It needs a mass flux of the order of $ 10^{-9} \; \mathrm{M}\dma{\earth} \; \mathrm{yr}^{-1} $
in order to be sustained.
This rate cannot be maintained through PR drag of material from the outer cold belt at about 140~AU,
but the inward scattering of small bodies by a chain of planets can marginally provide the required mass flux.

The hot ring seems to require an even higher replenishment rate
($ \sim 10^{-7} \; \mathrm{M}\dma{\earth} \; \mathrm{yr}^{-1} $).
It consists of small particles that are removed from the system by blowout.
A lower mass flux is possible if these grains are somehow retained near their production site,
lengthening their survival timescale.
The pile-up of dust due to the interplay of PR drag and sublimation is insufficient to explain the observations.
PR drag of dust grains from the warm belt could provide the required mass flux,
if a large amount of gas is present in the sublimation zone,
which slows down the blowout of unbound grains. 
The viability of such a substantial gas ring
and its consistency with existing observations remain to be tested.
Both oxygen and carbon will remain unaffected by radiation pressure around Fomalhaut.
In the case of the \bp\ debris disk, \citet{2006ApJ...643..509F} show that several species can potentially act as self-braking agents on the gas disk, and that more than 0.03 $M\dma{\oplus}$ of gas could be retained consistent with observed upper limits on the column densities.

%% file: conclusion.tex
\section{Discussion}\label{sec:discu}
In the last section, we showed that the hot exozodi of Fomalhaut could be the result of an accumulation of small unbound grains at the carbon sublimation distance brought there by the PR drag effect. 
For this mechanism to work, a trapping mechanism such as braking by a gaseous component needs to be invoked.
Alternative sources of continuum emission, such as free-free emission from a stellar wind, mass-loss events or hot gas, have been discussed \textit{e.g.} in Paper II, \citet{2008A&A...487.1041A} and \citet{2012A&A...546L...9D} and can be considered unsatisfying explanations: emission by very hot dust is the most convincing explanation to date. 
In the solar system, nanometer-sized particles are detected with the STEREO and Ulysse spacecrafts
\citep[][respectively]{2009SoPh..256..463M,2010P&SS...58..951K}, with an increase in the particle flux in the inner solar system. These very small grains, which are affected by the Lorentz force \citep{2010ApJ...714...89C}, could thus be expected in exozodiacal disks. In fact, magnetic trapping of nanograins could be a valuable alternative to the gas-braking mechanisms we have investigated in this study as discussed by \citet{2013ApJ...763..118S}. We note that it cannot be excluded that these nanograins might produce non-thermal emission, such as a PAH continuum. 

The bright hot component can be seen as the ``tip of the iceberg" in the sense that it may be a bright counterpart to the warm belt.
Our model of the warm belt confirms previous attempts to constrain its properties based on unresolved observations, although our detailed treatment of grain optics and the addition of spatial constraints point towards a closer in location than previously suspected. In particular using a blackbody model, \citet{2013ApJ...763..118S} estimated the belt location to be around 11 AU. 
While preserving the consistency with their dataset, the small FOV of the nulling interferometer impose the warm dust peak distance to be in the [1.5, 2.5] AU range after an appropriate subtraction of the hot dust contribution. 

Furthermore, future detection of a polarization signal could provide a confirmation of the model, and 
additional constraints on the grain properties and the disk geometry (in particular its inclination, assumed to match that of the cold belt in this study).
We use the MCFOST radiative transfer code \citep{2006A&A...459..797P,2009A&A...498..967P} with identical model parameters and assumptions as those discussed above\footnote{MCFOST uses a 3D geometry: we assume linear flaring, and we test scale heights of 0.001, 0.01 or 0.1 (unitless) at the reference distance of 1 AU, with little impact on the results. $r\dma{max}$ is fixed to 1\,AU and 20\,AU for the hot ring and the warm belt respectively. Here the grain temperatures are independent of their sizes.} in order to predict polarimetric signals for the exozodi. 
We find that the linear polarization integrated over the disk range from
$3\times10^{-7}$ to $5\times10^{-7}$ for the warm belt, and reaches 
$2\times10^{-5}$ to $3\times10^{-6}$ for the hot ring in bands U to I. 
These values are compatible with the upper limits of $9\times10^{-3}$ to $3\times10^{-3}$ given by \citet{2006A&A...452..921C}. 
For the hot ring, such signals would potentially be detectable by future sensitive polarimeters.

The warm belt has a distribution of temperatures that ranges from $\sim$320 to 470 K.
The suspected parent-body belt location is thus consistent with the prediction that such belts form preferentially before the snowline, $R\dma{snow} = 2.7\,(L/L\dma{\sun})^{1/2}\,\mathrm{AU} = 11\,\mathrm{AU}$, because giant planet accretion is triggered beyond that limit. 
At such distance, the parent-body population should have eroded in a short timescale compared to the age of the star. Given the properties of the cold dust belt, Poynting-Robertson drag is not a valuable mechanism to transport sufficient amount of material from $\sim$140 AU down to the warm belt. 
Even in a very favorable planetary system configuration, scattering of small bodies (comets, asteroids, planetesimals) from the cold belt by a chain of planets could not sustain the required mass production rate in the warm disk for the age of the star. 

Another explanation to the unexpectedly large mass in the warm component could be that the dust originates in stochastic and / or isolated catastrophic events, such as planetesimal collisions or break-up, or major dynamical perturbations. 
In the solar system, the Late Heavy Bombardment (LHB) was responsible for the depletion of the Kuiper-Belt, the release of large numbers of icy objects into its inner regions and probably a durable increase of the inner Zodiacal cloud infrared luminosity \citep{2010ApJ...713..816N,2009MNRAS.399..385B}.
The orbital parameters of Fomalhaut b have recently been reevaluated based on an fourth epoch detection with the HST. 
The planet orbit is found to be very eccentric ($\geq 0.8$) such that it likely approaches the innermost parts of the system at periastron, where additional planetary perturbers might be present \citep[][Beust et al., in prep.]{2013arXiv1305.2222K}.
Thus an LHB-like event may be occurring around Fomalhaut as a result of a planet-planet scattering event causing delayed stirring in both the cold and the warm belt.
In summary, a valuable scenario to understand the global debris disk is that a high collisional activity has been triggered by the presence of perturbing planets, reminiscent of the solar system history. 

Finally, it is attractive to place our study in the context of the long-term objective of finding and characterizing an Earth-like planet in the habitable zone of a star.
Scaling the Mars and Venus criteria for Fomalhaut \citep{2007A&A...476.1373S}, the habitable zone ranges between 2.5 and 5.5 AU. Under favorable 100\% cloud-cover conditions, it would extend from 1.4 to 8.1 AU. 
At these distances, the level of warm dust emission around Fomalhaut is high and represent therefore a threat for future spectroscopic and direct imaging missions \citep[e.g.,][]{2010A&A...509A...9D,2012PASP..124..799R}.
In turn, the existence of a massive asteroid belt may be an indication that there is no planet in these region as it would have cleared its neighborhood around its orbit.
A noticeable feature that we have not discussed yet is that the gap between the two exozodi components could be sculpted by the gravitational influence of a hidden planet at around 1 AU.
Constraints from radial velocities, astrometric measurements, and high-contrast imaging have been summarized in Paper I and are currently compatible with an hypothetical small mass companion in these regions.

\section{Conclusion}\label{sec:conclu}

In  a series of three papers, we have performed an interferometric study of the Fomalhaut inner debris disk. Paper I presented the detection of a circumstellar excess in K-band, attributed to very hot dust, confined well inside the $3$AU-HWHM FOV of the VINCI instrument. Despite the limited spatial constraints, the brightness temperature required calls for extremely hot, refractory grains lying very close to the star, at the sublimation limit.
Conversely, KIN null depth measurements presented in paper II are indicative of a warmer dust component at a few AU ($2$AU-HWHM FOV) that produces a rising excess upward of $10\mu$m. 
In the present study, we have presented the detailed results of self-consistent modeling of these two components by means of a parametric radiative transfer code and accurate treatment of debris disk physics.
To account for the expected size-dependent sublimation temperature of dust, we introduced a new prescription for the treatment of grain sublimation accounting for their specific dynamics and lifecycle.
This enabled us to assess realistically the spatial and size distribution of the emitting grains. 
We find that the Fomalhaut exozodiacal disk consists of two dust populations, one ``classical", though massive, disk of warm ($\sim 400$K) dust peaking at $\sim 2$AU and declining slowly with distance, responsible for most of the mid-infrared emission, and a hotter ($\sim$2000 K) and brighter counterpart dominated by small (0.01 - 0.5 $\mu$m), unbound dust particles at the limit of sublimation.
The stellar radius is approximately $9\times 10^{-3}$ AU and the hot grains are actually located at typically 10 to 35 stellar radii. 
The degeneracy inherent to SED fitting is partially broken by the spatial information contained in the interferometric data. We find that the model also fits the photometric mid/far-infrared measurements from Spitzer/MIPS and Herschel/PACS, and is consistent with the flux level measured in the Spitzer/IRS mid-infrared spectrum. If the warm dust, or an additional colder (but unresolved) component, were present further out in the system - as suggested by the suspected on-star excess from ALMA - it should produce moderate emission in the mid/far-infrared to preserve the compatibility between the KIN and Herschel / Spitzer data.

We analytically explored the various processes that can affect a dust grain: photo-gravitational and Poynting-Robertson drag forces, collisions, sublimation and disruption of big aggregates. We propose a framework for interpreting self-consistently the simultaneous prevalence of both hot and warm dust in the inner regions of Fomalhaut, similar to that also reported for samples of nearby main sequence stars by near- and mid- infrared exozodi surveys.
Firstly we find that neither of the two inner belts can be explained by a steady-state collisional cascade in a parent-body reservoir. 
Ignoring the production mechanism for the warm dust, we estimate that PR drag, from this component down to the sublimation radius, could transport enough mass into the hot component, provided that it can accumulate there. 
We showed that small carbon monomeres released by the disruption of larger aggregates that originate from the warm component can explain the observed flux level in the near-infrared, because this process considerably enhances the effective cross section of the dust population. 
Finally braking by a gaseous component could preserve these unbound grains from radiative transfer blowout for a sufficient time providing enough gas mass is available. If sublimation is the main source for this gas, it must have accumulated for a timescale comparable to the age of the star.

In summary, the intriguing hot dust phenomenon reported by various interferometric surveys could be understood in the light of the cumulation of multiple effects that eventually yield an accumulation of very small grains at a fraction of an AU. These hot rings are likely the counterparts of warm debris disks orbiting at a few AU that have their dust production triggered by intense collisional activity. In the near future, this scenario will need to be tested against statistical samples of objects, including later type stars.

%% file: rik4_app_black.tex
\section{The blowout timescale}\label{app:t_blow}

Particles with high $ \beta $ ratios are removed from the system by radiation pressure.
Here, we derive the typical timescale for this process,
for the case of $ \beta \gg 1 $.
In this limit, the transverse movement of the particles is small compared to the radial movement, and
only the radial acceleration of the particle needs to be considered.

\subsection{The gas free case}\label{app:t_blow_0}

We need to consider the forces of gravity and direct radiation pressure
(the PR drag force is negligibly small, and we do not consider gas drag at this stage).
These forces are given by
\begin{equation}
  \label{eq:f_rad}
  F\dma{rad} + F\dma{grav}
  = \frac{ ( \beta - 1 ) G M\dma{\star} m }{ r^2 }.
\end{equation}

For small radial displacements $ \Delta r $,
the acceleration $ \ddot{ r } = ( F\dma{rad} + F\dma{grav} ) / m $ can be assumed to be independent of $ r $,
and the displacement as a function of time is given by
$ \Delta r = \frac{ 1 }{ 2 } \ddot{ r } t^2 $.
The resulting timescale is
\begin{equation}
  \label{eq:t_dyn_small_dr}
  t\dma{ dyn , \, \Delta r \rightarrow 0 } ( \Delta r )
  = \sqrt{ \frac{ 2 \Delta r r\dma{release}^2 }{ G M\dma{\star} ( \beta - 1 ) } }.
\end{equation}

At large distances from the release point, the acceleration tends to zero,
and the velocity of the particle approaches a constant:
$ \dot{ r } ( r \rightarrow \infty ) = \sqrt{ 2 G M\dma{\star} \beta / r\dma{release} } $ (Lecavelier des Etangs et al. 1998).
Hence, for large displacements, the removal happens on a timescale of
\begin{equation}
  \label{eq:t_dyn_large_dr}
  t\dma{ dyn , \, \Delta r \rightarrow \infty } ( \Delta r )
  = \frac{ \Delta r }{ \dot{ r } ( r \rightarrow \infty ) }
  = \sqrt{ \frac{ ( \Delta r )^2 r\dma{release} }{ 2 G M\dma{\star} ( \beta - 1 ) } }.
\end{equation}


Adding Eqs.~\ref{eq:t_dyn_small_dr} and \ref{eq:t_dyn_large_dr} in quadrature
leads to the removal timescale
\begin{equation}
  \label{eq:t_dyn_comb}
  t\dma{dyn} ( \Delta r )
  = \sqrt{ \frac{ \Delta r r\dma{release} }{ G M\dma{\star} ( \beta - 1 ) }
    \left( 2 r\dma{release} + \frac{ \Delta r }{ 2 } \right) }.
\end{equation}
Comparing this equation with a numerical evaluation of the equation of motion shows that
its relative errors are less than 5\% for the values of $ \beta $ and $ r\dma{release} $ considered here.


We define $ t\dma{dyn} $ as the time it takes for a particle to fly
from its release point ($ r\dma{release} $)
to a point twice that radial distance from the star ($ 2 r\dma{release} $).
This is motivated by the fact that
the small grains are seen in a very narrow radial range,
so only the time they spend close to the release point is relevant.
Setting $ \Delta r = r\dma{release} = r $ gives the blowout timescale given by Eq.~\ref{eq:t_blow_0}.


\subsection{The high gas density case}\label{app:t_blow_rho}

For high gas densities, the gas drag force cannot be ignored.
Since the particles considered here are small (compared to the mean free path of the gas molecules),
and their velocities are low (compared to the sound speed of the gas),
the gas drag force is given by the subsonic Epstein drag law.
It is given by
\begin{equation}
  \label{eq:f_drag}
  F\dma{drag} = - \frac{ 4 \pi a^2 \rho\dma{g} v\dma{th} \Delta v }{ 3 },
\end{equation}
where $ \rho\dma{g} $ is the mass density of the gas, $ \Delta v $ is the relative speed between the dust grain and the gas,
and $ v\dma{th} $ is the mean thermal speed of the gas.
The latter is given by $ v\dma{th} = \sqrt{ 8 k\dma{B} T\dma{g} / ( \pi \mu\dma{g} m\dma{u} ) } $,
where $\mu\dma{g}$ is the molecular weight of gas molecules, $m\dma{u}$ is the atomic mass unit,
$k\dma{B}$ is the Boltzmann constant, and $T\dma{g}$ is the temperature of the gas.
To calculate $ v\dma{th} $, we assume
that the gas temperature equals the dust temperature ($T\dma{g} = T\dma{d}$),
and that the gas consists of the same molecules as the dust grains ($\mu\dma{g} = \mu\dma{d}$).

For high gas densities, the particles quickly reach the terminal velocity $ v\dma{term} $,
which is found by solving $ F\dma{rad} + F\dma{grav} + F\dma{drag} = 0 $ for $ \Delta v $.
This gives
\begin{equation}
  \label{eq:v_term}
  v\dma{term}
  = \frac{ G M\dma{\star} ( \beta - 1 ) \rho\dma{d} s }{ \rho\dma{g} v\dma{th} r^2 }.
\end{equation}
In the high gas density case, the average velocity
over the radial range $ \Delta r $
can be approximated by the terminal velocity.
The blowout timescale for high gas densities is then found from
$ t\dma{dyn, \, \rho\dma{g} \rightarrow \infty } ( \rho\dma{g}, \Delta r ) = \Delta r / v\dma{term} $,
which leads to Eq.~\ref{eq:t_blow_rho}.


%% file: bayes.tex
%
\begin{figure*}[h!tbp]\begin{center}  
\hbox to \textwidth{\includegraphics[angle=0,width=\textwidth,origin=bl]{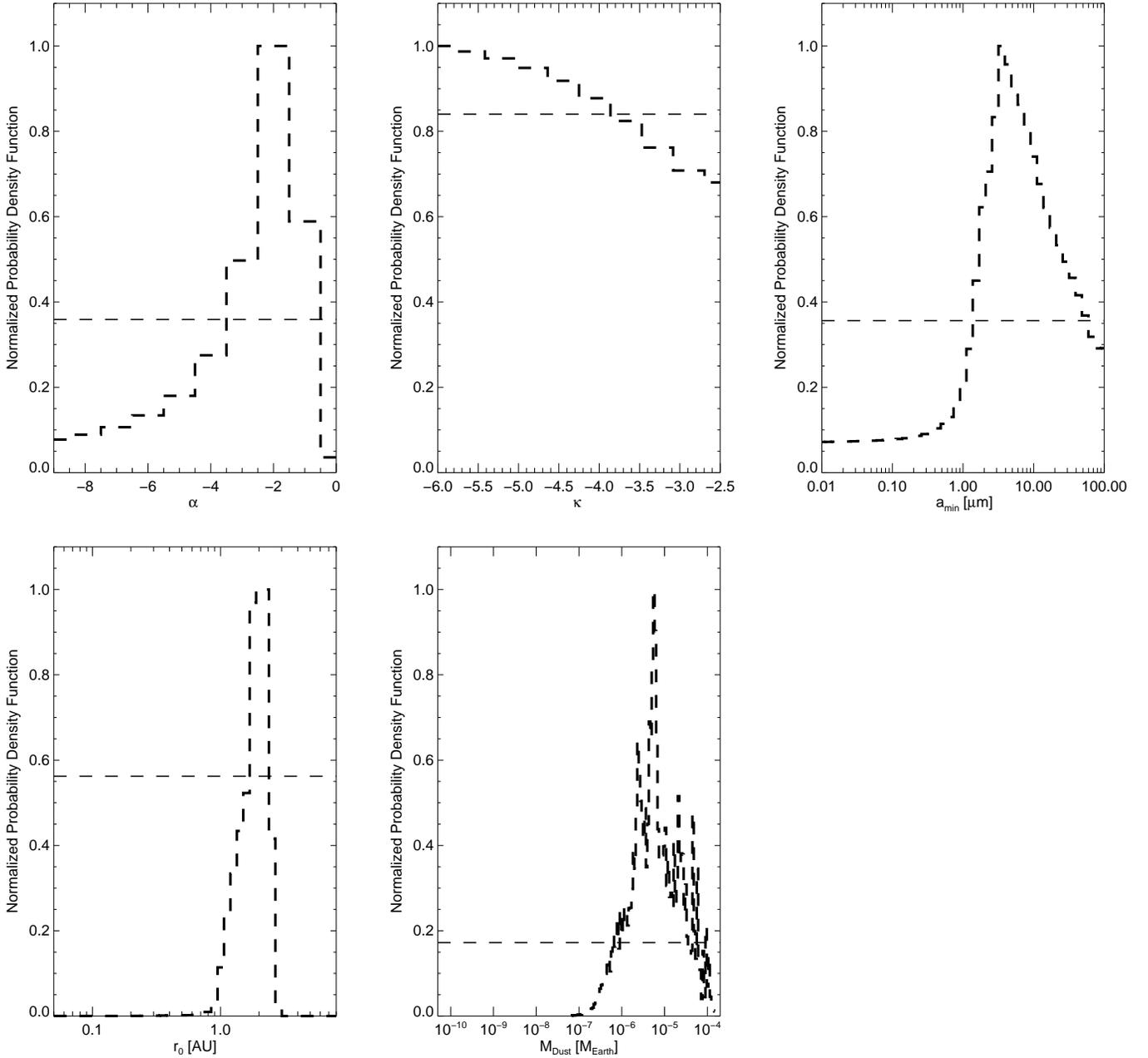}}
 \vspace{-0.3cm}
\caption{Bayesian probability curves obtained when fitting models to the warm component with approach 1 (inner density slope fixed to $10$, free outer density slope\,$\alpha$).}
 \end{center}\end{figure*}
 \vspace{-0.3cm}
\begin{figure*}[h!tbp]\begin{center}  
\hbox to \textwidth{\includegraphics[angle=0,width=\textwidth,origin=bl]{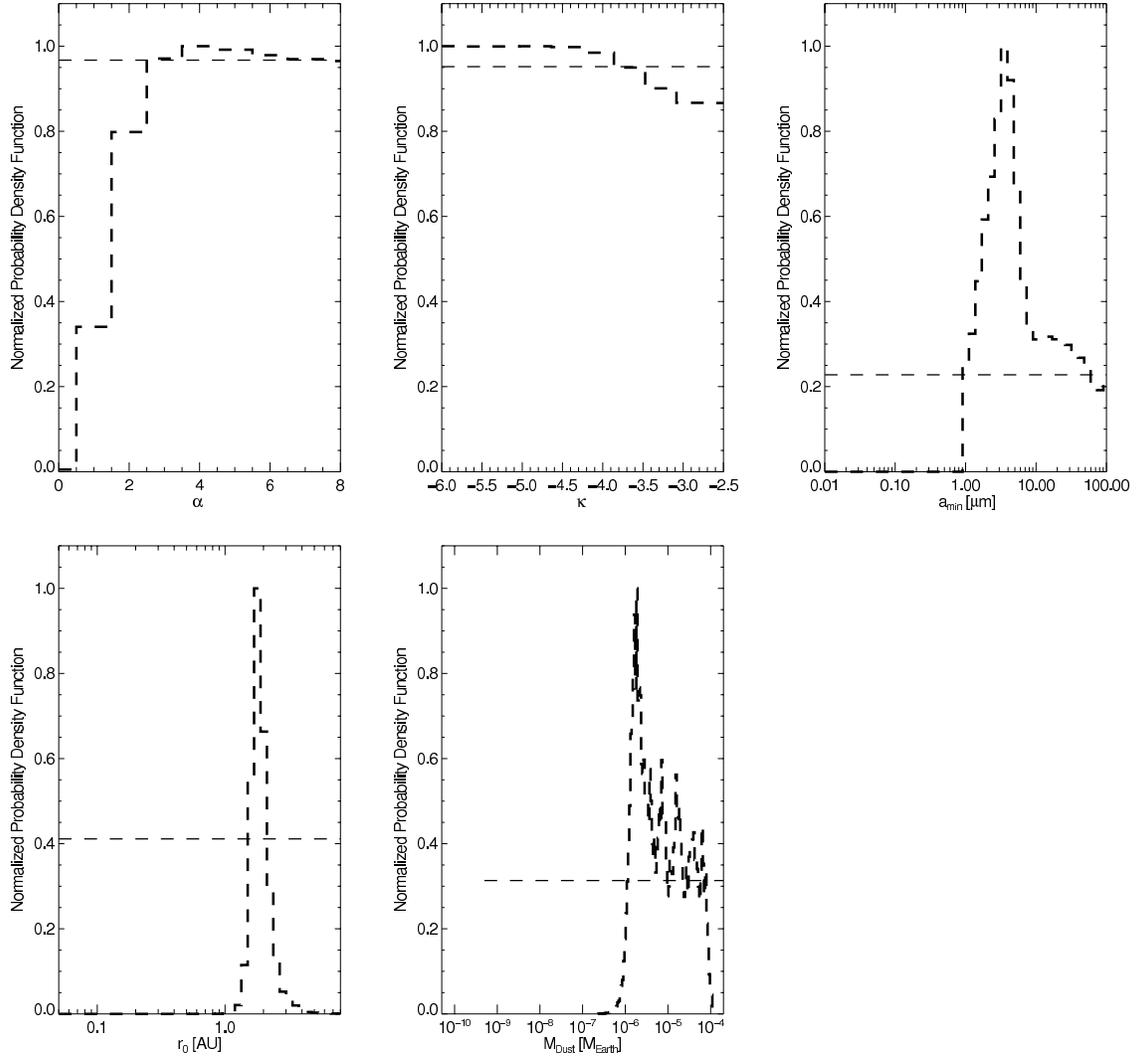}}
 \vspace{-0.3cm}
\caption{Bayesian probability curves obtained when fitting models to the warm component with approach 2 (outer density slope fixed to $1.5$, free inner density slope\,$\alpha$). The analysis uses some prior information regarding the grain sizes $P(a < a\dma{blow}/10 = 0$)}.
 \end{center}\end{figure*}